\documentclass[a4paper,reqno]{amsart}
\usepackage[centertags]{amsmath}

\usepackage{amssymb}
\usepackage{amsthm}
\usepackage{latexsym}
\theoremstyle{plain}

\newtheorem{proposition}{Proposition}
\newcommand{\al}{\alpha}
\newcommand{\be}{\beta}
\newcommand{\la}{\lambda}
\newcommand{\si}{\sigma}
\newcommand{\ro}{\rho}
\newcommand{\ti}[1]{\tilde{#1}}
\newcommand{\wti}[1]{\widetilde{#1}}
\newcommand{\pal}{{^{\scriptstyle{,}}\!\alpha}}
\newcommand{\pbe}{{^{\scriptstyle{,}}\!\beta}}
\newcommand{\pro}{{^{\scriptstyle{,}}\!\rho}}

\newcommand{\psig}{{^{\scriptstyle{,}}\!\sigma}}
\newcommand{\pmu}{{^{\scriptstyle{,}}\!\mu}}
\newcommand{\pnu}{{^{\scriptstyle{,}}\!\nu}}
\numberwithin{equation}{section}

\begin{document}

\title{Spin-2 gauge theories and perturbative gauge invariance}

\author{G\"unter Scharf}
\address{Institut f\"ur Theoretische Physik der Universit\"at Z\"urich \\
         Winterthurerstrasse 190  \\
         CH-8057 Z\"urich \\
         Switzerland}
\email{scharf@physik.unizh.ch}

\author{Mark Wellmann}
\address{Institut f\"ur Theoretische Physik der Universit\"at Z\"urich \\
         Winterthurerstrasse 190  \\
         CH-8057 Z\"urich \\
         Switzerland}
\email{wellmann@physik.unizh.ch}



\begin{abstract}
In the framework of causal perturbation theory we analyze the gauge structure of a massless self-interacting quantum tensor field. We look at this theory from a pure field theoretical point of view without assuming any geometrical aspect from general relativity. To first order in the perturbation expansion of the $S$-matrix we derive necessary and sufficient conditions for such a theory to be gauge invariant, by which we mean that the gauge variation of the self-coupling with respect to the gauge charge operator $Q$ is a divergence in the sense of vector analysis. The most general trilinear self-coupling of the graviton field turns out to be the one derived from the Einstein-Hilbert action plus divergences and coboundaries.      
\end{abstract}

\maketitle

\section{Introduction}
The general theory of relativity can be viewed as a theory for a self-interacting massless spin-$2$ field. This theory of gravity is derived from the Einstein-Hilbert (E-H) Lagrangian 
\begin{equation}
  L_{{\scriptscriptstyle EH}}=-\frac{2}{\kappa^2}\sqrt{-g}R
  \label{eq:Einstein-Hilbert-Lagrangefunktion}
\end{equation}
where $R=g^{\mu\nu}R_{\mu\nu}$ is the Ricci scalar and $\kappa^2=32\pi G$ ($G$ is Newton's gravitational constant). It is convenient to work with Goldberg variables~\cite{go:clgr}
\begin{equation}
  \ti{g}^{\mu\nu}=\sqrt{-g}g^{\mu\nu}
  \label{eq:Goldberg-Variablen}
\end{equation}
which one expands in an asymptotically flat geometry
\begin{equation}
  \ti{g}^{\mu\nu}=\eta^{\mu\nu}+\kappa h^{\mu\nu}
  \label{eq:Definition-Graviton-Feld}
\end{equation}
Here $\eta^{\mu\nu}$ is the metric of Minkowski spacetime. Then (\ref{eq:Einstein-Hilbert-Lagrangefunktion}) becomes an infinite power series in $\kappa$:
\begin{equation}
  L_{{\scriptscriptstyle EH}}=\sum_{j=0}^{\infty}\kappa^jL_{{\scriptscriptstyle EH}}^j
  \label{eq:Entwicklung-Einstein-Hilbert-Lagrangefunktion}
\end{equation}
The lowest order term $L_{{\scriptscriptstyle EH}}^0$ is quadratic in $h^{\mu\nu}(x)$ and defines the free asymptotic fields. In the Hilbert-gauge $h^{\mu\nu}_{\pmu}=0$, the graviton field $h^{\mu\nu}$ obeys the wave equation
\begin{equation}
  \Box h^{\mu\nu}(x)=0
  \label{eq:Wellengleichung-Graviton-Feld}
\end{equation}
The first order term $L_{{\scriptscriptstyle EH}}^1$ gives the trilinear coupling
\begin{equation}
  L_{{\scriptscriptstyle EH}}^1=\frac{1}{2}\,h^{\ro\si}\bigl(h^{\al\be}_{\pro}h^{\al\be}_{\psig}-\frac{1}{2}\,h^{}_{\pro}
                                h^{}_{\psig}+2\,h^{\al\ro}_{\pbe}h^{\be\si}_{\pal}+h^{}_{\pal}h^{\ro\si}_{\pal}
                                -2\,h^{\al\ro}_{\pbe}h^{\al\si}_{\pbe}\bigr)
  \label{eq:klassische-Einstein-Hilbert-Lagrangefunktion}
\end{equation}

There exists many alternative derivations of this result (\ref{eq:klassische-Einstein-Hilbert-Lagrangefunktion}), starting from massless tensor fields and requiring consistency with gauge invariance in some sense~\cite{fe:qtg,gu:qegfla,gu:eotg,ve:qtg,no:brstgf}. In classical theory the work closest to our non-geometrical point of view is the one of Ogievetsky and Polubarinov~\cite{op:ifs2ee}. These authors analyze spin-$2$ theories by working with a generalized Hilbert-gauge condition to exclude the spin one part from the outset. They impose an invariance under infinitesimal gauge transformations of the form
\begin{equation}
  \delta h^{\mu\nu}=\partial^{\mu}u^{\nu}+\partial^{\nu}u^{\mu}+\eta^{\mu\nu}\partial_{\al}u^{\al}
  \label{eq:klassische-Eichinvarianz}
\end{equation}
and get Einstein's theory at the end. Instead Wyss~\cite{wy:ug} and Deser~\cite{de:sgi} consider the coupling to matter. Then the self-coupling of the tensor-field (\ref{eq:klassische-Einstein-Hilbert-Lagrangefunktion}) is necessary for consistency. Wald~\cite{wa:stfgc} derives a divergence identity which is equivalent to an infinitesimal gauge invariance of the theory. Einstein's theory is the only non-trivial solution of this identity. In quantum theory the problem was studied by Boulware and Deser~\cite{bd:cgrqg}. These authors implement gauge invariance by requireing Ward identities associated with the graviton propagator. All authors get Einstein's theory as the unique classical limit if the theory is purely spin two without a spin one admixture.

In this paper we shall study the problem by means of perturbative quantum gauge invariance. This method which was worked out for spin-1 non-abelian gauge theories (massless~\cite{as:ym} and massive~\cite{ds:et}) in last years proceeds as follows: First one defines infinitesimal gauge variations on free fields. In the case of tensor fields it looks like (\ref{eq:klassische-Eichinvarianz}) where $u^{\mu}(x)$, instead of being an arbitrary function, is now a Fermi field which satisfies the wave equation. $u^{\mu}(x)$ may be regarded as a free Fadeev-Popov ghost field. Then we write down the most general trilinear coupling $T_1$ between the graviton and ghost fields which is compatible with Lorentz covariance, power counting and certain basic properties (like zero ghost number). Next we impose first order gauge invariance which strongly restricts the form of $T_1$. Among the possible solutions we recover Einstein's theory $L_{{\scriptscriptstyle EH}}^1$. The general solution can be written as a linear combination of $L_{{\scriptscriptstyle EH}}^1$ and divergences as well as coboundaries. In the perturbative construction of the $S$-matrix we next have to calculate the time-ordered product $T\{T_1(x)T_1(y)\}=T_2(x,y)$ by means of causality~\cite{eg:rlp,sch:qed}. Then Schorn~\cite{sch:giqgca} has shown that second order gauge invariance gives further restrictions, in particular, in the case of gravity it requires quartic normalisation terms of the form $L_{{\scriptscriptstyle EH}}^2$ in the above expansion (\ref{eq:Entwicklung-Einstein-Hilbert-Lagrangefunktion}). In this way the so-called proliferation of couplings can be overcome by perturbative gauge invariance.

The paper is organized as follows. In the next section we introduce perturbative gauge invariance. In section three we set up the general theory of a symmetrical tensor field by writing down all possible trilinear self-couplings and the most general ghost-coupling. In the following sections the consequences of first order gauge invariance are analysed. We obtain $15$ important conditions for spin-$2$ gauge theories. These conditions are necessary and sufficient for first order gauge invariance. 

\section{Perturbative gauge invariance}
Our fundamental free asymptotic fields are a symmetric tensor field of rank two $h^{\mu\nu}(x)$ and ghost and anti-ghost fields $u^{\mu}(x)$ and $\ti{u}^{\nu}(x)$. We consider these fields in the background of Minkowski spacetime. A symmetrical tensor field has ten degrees of freedom, which are more than the five independent components of a spin-$2$ field. The additional degrees of freedom can be eliminated by imposing two further conditions~\cite{op:ifs2ee}, namely
\begin{equation}
  h^{\mu\nu}(x)_{\pnu}=0\quad \text{and}\quad {h^{\mu}}_{\mu}(x)=0
  \label{eq:Elimination von Spin 0 und Spin 1 Komponenten}
\end{equation}
They are disregarded in the construction of the gauge theory and must be considered later in the characterization of physical states~\cite{gr:cqg}.

Our tensor field $h^{\mu\nu}(x)$ will be quantised as a massless field as follows
\begin{equation}
  \bigl[h^{\al\be}(x),h^{\mu\nu}(y)\bigr]=-ib^{\al\be\mu\nu}D_0(x-y)
  \label{eq:h-Quantisierung}
\end{equation}
where $D_0(x-y)$ is the massless Pauli-Jordan distribution and the tensor $b^{\al\be\mu\nu}$ is constructed from the Minkowski metric $\eta^{\mu\nu}$ in the following way
\begin{equation}
  b^{\al\be\mu\nu}=\frac{1}{2}\bigl(\eta^{\al\mu}\eta^{\be\nu}+\eta^{\al\nu}\eta^{\be\mu}-
                   \eta^{\al\be}\eta^{\mu\nu}\bigr)
  \label{eq:b-tensor}
\end{equation}
In analogy to spin-$1$ theories one introduces a gauge charge operator by
\begin{equation}
  Q:=\int\limits_{x^{0}=t} h^{\al\be}(x)_{\pbe}\overset{\leftrightarrow}{\partial_0}u^{\al}d^3x
  \label{eq:Q}
\end{equation}
For the construction of the physical subspace and in order to prove the unitarity of the $S$-matrix we want to have a nilpotent operator $Q$. Therefore we have to quantise the ghost fields with anticommutators
\begin{equation}
  \{u^{\mu}(x),\ti{u}^{\nu}(y)\}=i\eta^{\mu\nu}D_0(x-y)
  \label{eq:u-Quantisierung}
\end{equation}
and all other anti-commutators vanishing. All asymptotic fields fulfil the wave equation
\begin{equation}
  \begin{split}
    \Box\, h^{\mu\nu}(x)   & = 0 \\
    \Box\, u^{\al}(x)      & = 0 \\
    \Box\, \ti{u}^{\be}(x) & = 0 \\
  \end{split}
  \label{eq:Wellengleichung}
\end{equation}
The gauge charge $Q$ (\ref{eq:Q}) defines a gauge variation by
\begin{equation}
  d_QF:=QF-(-1)^{n_g(F)}FQ
  \label{eq:Eichvariation}
\end{equation}
where $n_g$ is the ghostnumber. This is the number of ghost fields minus the number of anti-ghost fields in the Wick monomial $F$. The operator $d_Q$ obeys the Leibniz rule
\begin{equation}
  d_Q(AB)=(d_QA)B+(-1)^{n_g(A)}Ad_QB
  \label{eq:Leibniz-Regel}
\end{equation}
where $A$ and $B$ are arbitrary operators.
We obtain the following gauge variations of the fundamental fields:
\begin{align}
  d_Qh^{\mu\nu}   & = -\frac{i}{2}\bigl(u^{\mu}_{\pnu}+u^{\nu}_{\pmu}-
                      \eta^{\mu\nu}u^{\al}_{\pal}\bigr) \label{eq:Eichvariation-h-Tensor} \\
  d_Qh            & = \ iu^{\mu}_{\pmu} \label{eq:Eichvariation-h-Spur} \\
  d_Q\ti{u}^{\mu} & = \ ih^{\mu\nu}_{\pnu} \label{eq:Eichvariation-Anti-Geist} \\
  d_Qu^{\mu}      & = \ 0 \label{eq:Eichvariation-Geist}
\end{align}
From (\ref{eq:Eichvariation-h-Tensor}) we immediately see
\begin{equation}
  d_Qh^{\mu\nu}_{\pmu}=0
  \label{eq:Eichvariation-h-Tensor-Ableitungskontraktion}
\end{equation}
The result (\ref{eq:Eichvariation-h-Tensor}) agrees with the infinitesimal gauge transformations of the Goldberg variables, so that our choice of $Q$ corresponds to the classical framework described in the introduction. The asymptotic fields will be used to construct the time-ordered products $T_n$ in the adiabatically switched $S$-matrix
\begin{equation}
  S(g)=\mathbf{1}+\sum_{n=1}^{\infty}\frac{1}{n!}\int T_n(x_1,\ldots,x_n)g(x_1)\ldots g(x_n)\,d^4x_1\ldots d^4x_n
  \label{eq:S-Matrix}
\end{equation}
where $g\in\mathcal{S}(\mathbb{R}^4)$ is a test function. The time ordered products $T_n$ are operator valued distributions and can be expressed by normally ordered products of free fields. It is very important that gauge invariance of the $S$-matrix can be directly formulated in terms of the $T_n$. First order gauge invariance means that $d_QT_1$ is a divergence in the sense of vector analysis, i.e.
\begin{equation}
  d_QT_1(x)=i\partial_{\mu}T_{1/1}^{\mu}(x)
  \label{eq:Eichinvarianz-1.-Ordnung}
\end{equation}
The definition of the $n$-th order gauge invariance then reads 
\begin{equation}
  d_QT_n=\bigl[Q,T_n\bigr]=i\sum_{l=1}^{n}\frac{\partial}{\partial x^{\mu}_l}T_{n/l}^{\mu}(x_1,\ldots,x_l,\ldots,x_n)
  \label{eq:Eichinvarianz-n.te-Ordnung}
\end{equation}
Here $T_{n/l}^{\mu}$ is the time ordered product with a gauge variated vertex $T_{1/1}^{\mu}(x_l)$ at position $x_l$ and ordinary vertices $T_1$ at the other arguments.  
 
\section{Structure of the Interaction}
Here we introduce the self-couplings of the quantum tensor field. We consider for this purpose a symmetrical rank-$2$ tensor field in a fixed background, namely Minkowski spacetime. The simplest expression leading to a self-interacting spin-$2$ field theory is a trilinear coupling of the quantum fields $h^{\mu\nu}(x)$ and $h(x)\equiv {h^{\mu}}_{\mu}(x)$. We require Lorentz invariance and in addition to that two derivatives acting on the fields. This is for the following reasons: First of all, by inspection of all trilinear self-interaction terms without derivatives, it is easily seen that such a theory cannot be gauge invariant to first order of perturbation theory. Therefore an interaction without derivatives can be ruled out. Secondly it is impossible to form a Lorentz-scalar from three rank-$2$ tensor fields with only one derivative. Last but not least the corresponding trilinear expression in the expansion the E-H action contains two derivatives as well. Therefore we are able to reproduce the results from classical general relativity.

In the following all fields are free fields obeying the free field equations of motion. All products of two or more fields at the same spacetime point $x$ are viewed as normal products. Then the general ansatz for a combination of three field operators contains $12$ terms\footnote{We use the following convention regarding the indices. All vector- and tensor indices are written as superscript, whereas all partial derivatives are written as subscript in the abbreviated form with a prime in front of the index, i.e.: $A(x)_{\pnu}=\partial A(x)/\partial x^{\nu}$. All indices will be raised and lowered by the Minkowski-metric $\eta_{\mu\nu}$ and will be properly contracted like $A^{\mu}B^{\mu}:=\eta_{\mu\nu}A^{\mu}B^{\nu}$.}:
\begin{equation}
  \begin{split}
    T_1^h(x):= & \ a_1:h^{\mu\nu}(x)_{\pmu}h^{}(x)_{\pnu}h(x):+\,a_2:h^{\mu\nu}(x)h^{}(x)_{\pmu}
                 h^{}(x)_{\pnu}: \\
               & +a_3:h^{\al\be}(x)_{\pal}h^{\be\mu}(x)_{\pmu}h(x):+\,a_4:h^{\al\be}(x)_{\pal}h^{\be\mu}(x)
                 h^{}(x)_{\pmu}: \\
               & +a_5:h^{\al\be}(x)h^{\be\mu}(x)_{\pal}h^{}(x)_{\pmu}:+\,a_6:h^{\al\be}(x)_{\pmu}
                 h^{\be\mu}(x)_{\pal}h(x): \\
               & +a_7:h^{\mu\nu}(x)_{\pmu}h^{\al\be}(x)_{\pnu}h^{\al\be}(x):+\,a_8:h^{\mu\nu}(x)
                 h^{\al\be}(x)_{\pmu}h^{\al\be}(x)_{\pnu}: \\
               & +a_9:h^{\mu\nu}(x)_{\pal}h^{\nu\al}(x)_{\pbe}h^{\mu\be}(x):+\,a_{10}:h^{\mu\nu}(x)_{\pal}
                 h^{\nu\al}(x)h^{\mu\be}(x)_{\pbe}: \\
               & +a_{11}:h^{\mu\nu}(x)h^{\nu\al}(x)_{\pal}h^{\mu\be}(x)_{\pbe}:+\,a_{12}:h^{\mu\nu}(x)
                 h^{\nu\al}(x)_{\pbe}h^{\mu\be}(x)_{\pal}: \\
  \end{split}
  \label{eq:T1h}
\end{equation}
Here we have omitted all terms which are gauge invariant in a trivial way. These are terms with a contraction on the two derivatives, e.g. $h^{}(x)_{\pal}h^{}(x)_{\pal}h(x)=1/2\,\partial_{\al}\bigl(h^{}(x)_{\pal}h(x)h(x)\bigr)$. Furthermore all terms with two derivatives acting on the same field can be transformed into a divergence plus a term already contained in (\ref{eq:T1h}). These terms would modify our ansatz only in a redefinition of some parameters $a_i$ and can be omitted without losing generality.

As in the cases of Yang-Mills theory~\cite{dhks:ccymt1,dhks:ccymt2} and Einstein gravity~\cite{sch:giqgca} we expect to get a gauge invariant first order coupling only if we couple the tensor field $h^{\mu\nu}$ to ghost and anti-ghost fields. The most general expression with zero ghost-number is 
\begin{equation}
  \begin{split} 
    T_1^u(x):= & \ b_1:u^{\ro}(x)_{\pnu}\ti{u}^{\mu}(x)_{\pro}h^{\mu\nu}(x):+\,b_2:u^{\ro}(x)_{\pnu}
                 \ti{u}^{\mu}(x)h^{\mu\nu}(x)_{\pro}: \\
               & +b_3:u^{\ro}(x)\ti{u}^{\mu}(x)_{\pnu}h^{\mu\nu}(x)_{\pro}:+\,b_4:u^{\ro}(x)_{\pro}
                 \ti{u}^{\mu}(x)_{\pnu}h^{\mu\nu}(x): \\
               & +b_5:u^{\ro}(x)_{\pro}\ti{u}^{\mu}(x)h^{\mu\nu}(x)_{\pnu}:+\,b_6:u^{\ro}(x)
                 \ti{u}^{\mu}(x)_{\pro}h^{\mu\nu}(x)_{\pnu}: \\
               & +b_7:u^{\ro}(x)_{\pmu}\ti{u}^{\mu}(x)_{\pro}h(x):+\,b_8:u^{\ro}(x)_{\pmu}\ti{u}^{\mu}(x)
                 h^{}(x)_{\pro}: \\
               & +b_9:u^{\ro}(x)\ti{u}^{\mu}(x)_{\pmu}h^{}(x)_{\pro}:+\,b_{10}:u^{\ro}(x)_{\pro}
                 \ti{u}^{\mu}(x)_{\pmu}h(x): \\
               & +b_{11}:u^{\ro}(x)_{\pro}\ti{u}^{\mu}(x)h^{}(x)_{\pmu}:+\,b_{12}:u^{\ro}(x)
                 \ti{u}^{\mu}(x)_{\pro}h^{}(x)_{\pmu}: \\
               & +b_{13}:u^{\ro}(x)_{\pmu}\ti{u}^{\mu}(x)_{\pnu}h^{\ro\nu}(x):+\,b_{14}:u^{\ro}(x)_{\pmu}
                 \ti{u}^{\mu}(x)h^{\ro\nu}(x)_{\pnu}: \\
               & +b_{15}:u^{\ro}(x)\ti{u}^{\mu}(x)_{\pmu}h^{\ro\nu}(x)_{\pnu}:+\,b_{16}:u^{\ro}(x)_{\pnu}
                 \ti{u}^{\mu}(x)_{\pmu}h^{\ro\nu}(x): \\
               & +b_{17}:u^{\ro}(x)_{\pnu}\ti{u}^{\mu}(x)h^{\ro\nu}(x)_{\pmu}:+\,b_{18}:u^{\ro}(x)
                 \ti{u}^{\mu}(x)_{\pnu}h^{\ro\nu}(x)_{\pmu}: \\
               & +b_{19}:u^{\mu}(x)_{\pnu}\ti{u}^{\mu}(x)_{\pro}h^{\ro\nu}(x):+\,b_{20}:u^{\mu}(x)_{\pnu}
                 \ti{u}^{\mu}(x)h^{\ro\nu}(x)_{\pro}: \\
               & +b_{21}:u^{\mu}(x)\ti{u}^{\mu}(x)_{\pnu}h^{\ro\nu}(x)_{\pro}: \\
  \end{split}
  \label{eq:T1u}
\end{equation} 
We will suppress all arguments of the field operators as well as the double dots of normal ordering in subsequent expressions. The complete first order coupling is then given by:
\begin{equation}
  T_1 := T_1^h+T_1^u
  \label{eq:T1}
\end{equation}
In the following analysis we are interested in the question how the parameters of the theory $a_1,\ldots, a_{12}$ and $b_1,\ldots, b_{21}$ will be restricted due to first order gauge invariance.

\section{Gauge invariance to first order}

\subsection{Ansatz for a divergence}
In the previous section we have defined our trilinear coupling $T_1^h$ as well as the coupling to ghost- and anti-ghost fields $T_1^u$. In this section we try to write the gauge variation $d_QT_1$ as a divergence $\partial_{\mu}T_{1/1}^{\mu}$. We proceed in the following way: Because of the great variety of different terms in $d_QT_1$ it is most convenient to use a separate ansatz for $T_{1/1}^{\mu}$. Since the operator $d_Q$ applied to our $T_1$ increases the ghostnumber of the result by one we have to make an ansatz with $n_g(T_{1/1}^{\mu})=1$. The terms appearing in this ansatz can be classified according to their index structure regarding the tensor indices: There are seven different types of the form $uhh$, namely

1.) Type $A$:
\begin{equation}
   \begin{split}
     T_{1/1}^{\mu,A} = & \ c_1\, u^{\mu}_{\pal} h^{\ro\si}_{\pal} h^{\ro\si}+
                         c_2\, u^{\mu} h^{\ro\si}_{\pal} h^{\ro\si}_{\pal}+
                         c_3\, u^{\al}_{\pal\pmu} h^{\ro\si} h^{\ro\si}+
                         c_4\, u^{\al} h^{\ro\si}_{\pal\pmu} h^{\ro\si} \\
                       & +c_5\, u^{\al}_{\pal} h^{\ro\si}_{\pmu} h^{\ro\si}+
                         c_6\, u^{\al} h^{\ro\si}_{\pal} h^{\ro\si}_{\pmu}+
                         c_7\, u^{\al}_{\pmu} h^{\ro\si}_{\pal} h^{\ro\si} \\
   \end{split}
  \label{eq:TypA}
\end{equation}

2.) Type $B$:
\begin{equation}
   \begin{split}
     T_{1/1}^{\mu,B} = & \ c_8\, u^{\mu}_{\pal} h^{}_{\pal} h+
                           c_9\, u^{\mu} h^{}_{\pal} h^{}_{\pal}+
                           c_{10}\, u^{\al}_{\pal\pmu} h h+ 
                           c_{11}\, u^{\al} h^{}_{\pal\pmu} h \\ 
                       & + c_{12}\, u^{\al}_{\pal} h^{}_{\pmu} h+ 
                           c_{13}\, u^{\al} h^{}_{\pal} h^{}_{\pmu}+
                           c_{14}\, u^{\al}_{\pmu} h_{\pal} h \\
   \end{split}
  \label{eq:TypB}
\end{equation}

3.) Type $C$:
\begin{equation}
   \begin{split}
     T_{1/1}^{\mu,C} = & \ c_{15}\, u^{\al}_{\pnu} h^{\al\mu}_{\pnu} h+
                           c_{16}\, u^{\al}_{\pnu} h^{\al\mu} h^{}_{\pnu}+
                           c_{17}\, u^{\al} h^{\al\mu}_{\pnu} h^{}_{\pnu}+
                           c_{18}\, u^{\al}_{\pnu\pmu} h^{\al\nu} h \\
                       & + c_{19}\, u^{\al} h^{\al\nu}_{\pnu\pmu} h+
                           c_{20}\, u^{\al} h^{\al\nu} h^{}_{\pnu\pmu}+ 
                           c_{21}\, u^{\al}_{\pnu} h^{\al\nu}_{\pmu} h+
                           c_{22}\, u^{\al}_{\pnu} h^{\al\nu} h^{}_{\pmu} \\
                       & + c_{23}\, u^{\al} h^{\al\nu}_{\pnu} h^{}_{\pmu}+ 
                           c_{24}\, u^{\al}_{\pmu} h^{\al\nu}_{\pnu} h+
                           c_{25}\, u^{\al}_{\pmu} h^{\al\nu} h^{}_{\pnu}+
                           c_{26}\, u^{\al} h^{\al\nu}_{\pmu} h^{}_{\pnu} \\
   \end{split}
  \label{eq:TypC}
\end{equation}

4.) Type $D$:
\begin{equation}
   \begin{split}
     T_{1/1}^{\mu,D} = & \ c_{27}\, u^{\al}_{\pro} h^{\al\si}_{\pro} h^{\si\mu}+ 
                           c_{28}\, u^{\al}_{\pro} h^{\al\si} h^{\si\mu}_{\pro}+
                           c_{29}\, u^{\al} h^{\al\si}_{\pro} h^{\si\mu}_{\pro}+
                           c_{30}\, u^{\al}_{\pro\pmu} h^{\al\si} h^{\si\ro} \\
                       & + c_{31}\, u^{\al} h^{\al\si}_{\pro\pmu} h^{\si\ro}+
                           c_{32}\, u^{\al} h^{\al\si} h^{\si\ro}_{\pro\pmu}+
                           c_{33}\, u^{\al}_{\pro} h^{\al\si}_{\pmu} h^{\si\ro}+
                           c_{34}\, u^{\al}_{\pro} h^{\al\si} h^{\si\ro}_{\pmu} \\ 
                       & + c_{35}\, u^{\al} h^{\al\si}_{\pro} h^{\si\ro}_{\pmu}+
                           c_{36}\, u^{\al}_{\pmu} h^{\al\si}_{\pro} h^{\si\ro}+
                           c_{37}\, u^{\al}_{\pmu} h^{\al\si} h^{\si\ro}_{\pro}+
                           c_{38}\, u^{\al} h^{\al\si}_{\pmu} h^{\si\ro}_{\pro} \\
   \end{split}
  \label{eq:TypD}
\end{equation}

5.) Type $E$:
\begin{equation}
   \begin{split}
     T_{1/1}^{\mu,E} = & \ c_{39}\, u^{\al}_{\psig\pro} h^{\al\ro} h^{\mu\si}+
                           c_{40}\, u^{\al} h^{\al\ro}_{\psig\pro} h^{\mu\si}+
                           c_{41}\, u^{\al} h^{\al\ro} h^{\mu\si}_{\psig\pro}+
                           c_{42}\, u^{\al}_{\psig} h^{\al\ro}_{\pro} h^{\mu\si} \\
                       & + c_{43}\, u^{\al}_{\psig} h^{\al\ro} h^{\mu\si}_{\pro}+
                           c_{44}\, u^{\al} h^{\al\ro}_{\psig} h^{\mu\si}_{\pro}+
                           c_{45}\, u^{\al}_{\pro} h^{\al\ro}_{\psig} h^{\mu\si}+
                           c_{46}\, u^{\al}_{\pro} h^{\al\ro} h^{\mu\si}_{\psig} \\
                       & + c_{47}\, u^{\al} h^{\al\ro}_{\pro} h^{\mu\si}_{\psig}+
                           c_{48}\, u^{\al}_{\pro\psig} h^{\al\mu} h^{\si\ro}+
                           c_{49}\, u^{\al} h^{\al\mu}_{\pro\psig} h^{\ro\si}+
                           c_{50}\, u^{\al} h^{\al\mu} h^{\ro\si}_{\pro\psig} \\
                       & + c_{51}\, u^{\al}_{\pro} h^{\al\mu}_{\psig} h^{\ro\si}+
                           c_{52}\, u^{\al}_{\pro} h^{\al\mu} h^{\ro\si}_{\psig}+
                           c_{53}\, u^{\al} h^{\al\mu}_{\pro} h^{\ro\si}_{\psig}\\
   \end{split}
  \label{eq:TypE}
\end{equation}

6.) Type $F$:
\begin{equation}
   \begin{split}
     T_{1/1}^{\mu,F} = & \ c_{54}\, u^{\ro}_{\psig\pro} h^{\mu\nu} h^{\nu\si}+ 
                           c_{55}\, u^{\ro} h^{\mu\nu}_{\psig\pro} h^{\nu\si}+
                           c_{56}\, u^{\ro} h^{\mu\nu} h^{\nu\si}_{\psig\pro}+
                           c_{57}\, u^{\ro}_{\psig} h^{\mu\nu}_{\pro} h^{\nu\si} \\
                       & + c_{58}\, u^{\ro}_{\psig} h^{\mu\nu} h^{\nu\si}_{\pro}+ 
                           c_{59}\, u^{\ro} h^{\mu\nu}_{\psig} h^{\nu\si}_{\pro}+
                           c_{60}\, u^{\ro}_{\pro} h^{\mu\nu}_{\psig} h^{\nu\si}+
                           c_{61}\, u^{\ro}_{\pro} h^{\mu\nu} h^{\nu\si}_{\psig} \\
                       & + c_{62}\, u^{\ro} h^{\mu\nu}_{\pro} h^{\nu\si}_{\psig}+ 
                           c_{63}\, u^{\mu}_{\pro\psig} h^{\ro\nu} h^{\nu\si}+
                           c_{64}\, u^{\mu} h^{\ro\nu}_{\pro\psig} h^{\nu\si}+
                           c_{65}\, u^{\mu} h^{\ro\nu}_{\psig} h^{\nu\si}_{\pro} \\
                       & + c_{66}\, u^{\mu}_{\pro} h^{\ro\nu}_{\psig} h^{\nu\si}+ 
                           c_{67}\, u^{\mu}_{\pro} h^{\ro\nu} h^{\nu\si}_{\psig}+
                           c_{68}\, u^{\mu} h^{\ro\nu}_{\pro} h^{\nu\si}_{\psig} \\
   \end{split}
  \label{eq:TypF}
\end{equation}

7.) Type $G$:
\begin{equation}
   \begin{split}
     T_{1/1}^{\mu,G} = & \ c_{69}\, u^{\ro}_{\psig\pro} h^{\mu\si} h+
                           c_{70}\, u^{\ro} h^{\mu\si}_{\psig\pro} h+ 
                           c_{71}\, u^{\ro} h^{\mu\si} h^{}_{\psig\pro}+
                           c_{72}\, u^{\ro}_{\psig} h^{\mu\si}_{\pro} h \\
                       & + c_{73}\, u^{\ro}_{\psig} h^{\mu\si} h^{}_{\pro}+
                           c_{74}\, u^{\ro} h^{\mu\si}_{\psig} h^{}_{\pro}+
                           c_{75}\, u^{\ro}_{\pro} h^{\mu\si}_{\psig} h+
                           c_{76}\, u^{\ro}_{\pro} h^{\mu\si} h^{}_{\psig} \\
                       & + c_{77}\, u^{\ro} h^{\mu\si}_{\pro} h^{}_{\psig}+
                           c_{78}\, u^{\mu}_{\pro\psig} h^{\ro\si} h+ 
                           c_{79}\, u^{\mu} h^{\ro\si}_{\pro\psig} h+
                           c_{80}\, u^{\mu} h^{\ro\si} h^{}_{\pro\psig} \\
                       & + c_{81}\, u^{\mu}_{\pro} h^{\ro\si}_{\psig} h+
                           c_{82}\, u^{\mu}_{\pro} h^{\ro\si} h^{}_{\psig}+
                           c_{83}\, u^{\mu} h^{\ro\si}_{\pro} h^{}_{\psig} \\
   \end{split}
  \label{TypG}
\end{equation}
The remaining ones are products of two ghost fields and one anti-ghost field. Here we have three different types:
 
8.) Type $H$:
\begin{equation}
   \begin{split}
     T_{1/1}^{\mu,H} = & \ c_{84}\, u^{\mu}_{\psig} \ti{u}^{\al}_{\psig} u^{\al}+
                           c_{85}\, u^{\mu}_{\psig} \ti{u}^{\al} u^{\al}_{\psig}+
                           c_{86}\, u^{\mu} \ti{u}^{\al}_{\psig} u^{\al}_{\psig}+ 
                           c_{87}\, u^{\si}_{\psig} \ti{u}^{\al}_{\pmu} u^{\al} \\
                       & + c_{88}\, u^{\si}_{\psig} \ti{u}^{\al} u^{\al}_{\pmu}+
                           c_{89}\, u^{\si} \ti{u}^{\al}_{\psig} u^{\al}_{\pmu}+
                           c_{90}\, u^{\si}_{\pmu} \ti{u}^{\al}_{\psig} u^{\al}+ 
                           c_{91}\, u^{\si}_{\pmu} \ti{u}^{\al} u^{\al}_{\psig} \\
                       & + c_{92}\, u^{\si} \ti{u}^{\al}_{\pmu} u^{\al}_{\psig}+
                           c_{93}\, u^{\si}_{\psig\pmu} \ti{u}^{\al} u^{\al}+
                           c_{94}\, u^{\si} \ti{u}^{\al}_{\psig\pmu} u^{\al}+
                           c_{95}\, u^{\si} \ti{u}^{\al} u^{\al}_{\psig\pmu} \\
   \end{split}
  \label{eq:TypH}
\end{equation}

9.) Type $J$:
\begin{equation}
   \begin{split}
     T_{1/1}^{\mu,J} = & \ c_{96}\, u^{\al}_{\pro} \ti{u}^{\mu}_{\pal} u^{\ro}+
                           c_{97}\, u^{\al}_{\pal} \ti{u}^{\mu}_{\pro} u^{\ro}+
                           c_{98}\, u^{\al} \ti{u}^{\mu}_{\pro\pal} u^{\ro}+
                           c_{99}\, u^{\al}_{\pro\pal} \ti{u}^{\mu} u^{\ro} \\
                       & + c_{100}\, u^{\ro}_{\pal\pro} \ti{u}^{\al} u^{\mu}+
                           c_{101}\, u^{\ro} \ti{u}^{\al}_{\pal\pro} u^{\mu}+ 
                           c_{102}\, u^{\ro} \ti{u}^{\al} u^{\mu}_{\pal\pro}+
                           c_{103}\, u^{\ro}_{\pal} \ti{u}^{\al}_{\pro} u^{\mu} \\
                       & + c_{104}\, u^{\ro}_{\pal} \ti{u}^{\al} u^{\mu}_{\pro}+
                           c_{105}\, u^{\ro} \ti{u}^{\al}_{\pal} u^{\mu}_{\pro}+ 
                           c_{106}\, u^{\ro}_{\pro} \ti{u}^{\al}_{\pal} u^{\mu}+
                           c_{107}\, u^{\ro}_{\pro} \ti{u}^{\al} u^{\mu}_{\pal} \\
                       & + c_{108}\, u^{\ro} \ti{u}^{\al}_{\pro} u^{\mu}_{\pal}+
                           c_{109}\, u^{\mu} \ti{u}^{\al}_{\pal\pro} u^{\ro} \\
   \end{split}
  \label{eq:TypJ}
\end{equation}
 
10.) Type $K$:
\begin{equation}
   \begin{split}
     T_{1/1}^{\mu,K} = & \ c_{110}\, u^{\si}_{\pal} \ti{u}^{\mu}_{\pal} u^{\si}+
                           c_{111}\, u^{\si}_{\pal\pmu} \ti{u}^{\al} u^{\si}+
                           c_{112}\, u^{\si}_{\pal} \ti{u}^{\al}_{\pmu} u^{\si} \\
                       & + c_{113}\, u^{\si}_{\pal} \ti{u}^{\al} u^{\si}_{\pmu}+
                           c_{114}\, u^{\si}_{\pmu} \ti{u}^{\al}_{\pal} u^{\si} \\ 
   \end{split}
  \label{eq:TypK}
\end{equation}  
Then we obtain the total divergence as the sum of these 10 different types
\begin{equation}
  \partial_{\mu} T_{1/1}^{\mu} = \partial_{\mu}\sum_{i\in \{A,\ldots, K\}}T_{1/1}^{\mu,i}
  \label{eq:T1/1}
\end{equation}
The parameters $c_1,\ldots,c_{114}\in\mathbb{C}$ are for the moment free constants, to be determined by gauge invariance. This expression\footnote{$T_{1/1}^{\mu}$ is called $Q$-vertex in the sequel because it is obtained from the usual vertex $T_1$ if one replaces a quantum field with the gauge variation of that field.} for $T_{1/1}^{\mu}$ contains all possible combinations of fields appearing after gauge variation of $T_1$. Without losing generality one can now eliminate a few terms in the types $A,\ldots,D,H$ and $K$\footnote{This relies on an idea of M.D\"utsch, see \cite{due:nqym}}. Therefore we consider a new $Q$-vertex $\wti{T}_{1/1}^{\mu}(x)$ for which the following relation holds
\begin{equation}
  T_{1/1}^{\mu}(x) = \wti{T}_{1/1}^{\mu}(x) + B^{\mu}(x)
  \label{eq:tiT1/1}
\end{equation}
where $B^{\mu}$ has the special form $B^{\mu}(x)=\partial_{\nu}^xA^{\nu\mu}(x)$ and $A^{\nu\mu}(x)$ is an anti-symmetrical tensor of rank $2$. Then we have
\begin{equation}
  \partial_{\mu} T_{1/1}^{\mu}(x) = \partial_{\mu} \wti{T}_{1/1}^{\mu}(x),
  \label{eq:T1/1=tiT1/1}
\end{equation}
because partial derivatives are commuting. Let us now construct such a tensor $A^{\nu\mu}$. We consider the type-$A$ term $c_3\,u^\al_{\pal\pmu}h^{\ro\si}h^{\ro\si}$. This can be written as
\begin{equation}
  c_3\,u^\al_{\pal\pmu}h^{\ro\si}h^{\ro\si} =
  c_3\,\bigl[\partial_{\al}\bigl(u^{\al}_{\pmu}h^{\ro\si}h^{\ro\si}\bigr) -
  2\,u^{\al}_{\pmu}h^{\ro\si}_{\pal}h^{\ro\si}\bigr]
  \label{eq:c3-Term}
\end{equation}
In an analogous way and using the wave equation we can write 
\begin{equation}
  0 = c_3\,u^{\mu}_{\pal\pal}h^{\ro\si}h^{\ro\si} =
  c_3\,\bigl[\partial_{\al}\bigl(u^{\mu}_{\pal}h^{\ro\si}h^{\ro\si}\bigr) -
  2\,u^{\mu}_{\pal}h^{\ro\si}_{\pal}h^{\ro\si}\bigr]
  \label{eq:c3-Term=0}
\end{equation}
Now we add $-c_3\,u^{\mu}_{\pal\pal}h^{\ro\si}h^{\ro\si}$ to $T_{1/1}^{\mu}$ and obtain
\begin{equation}
  T_{1/1}^{\mu} = \wti{T}_{1/1}^{\mu} +
  c_3\,\partial_{\nu}\bigl(u^{\nu}_{\pmu} - u^{\mu}_{\pnu}\bigr)h^{\ro\si}h^{\ro\si}
  \label{eq:tiT1/1-Beispiel}
\end{equation}
The expression in brackets is anti-symmetric in $\nu,\mu$ and we get $\wti{T}_{1/1}^{\mu}$ if we replace the constants $c_1$ with $c_1+2\,c_3$ and $c_7$ with $c_7-2\,c_3$ in $T_{1/1}^{\mu}$. In this way we can eliminate the monomials with constants $c_3,c_4$ in type $A$, $c_{10},c_{11}$ in type $B$, $c_{18},c_{19},c_{20}$ in type $C$, $c_{30},c_{31},c_{32}$ in type $D$, $c_{93},c_{94},c_{95}$ in type $H$ and $c_{111}$ in type $K$. Then we obtain a smaller $Q$-vertex $\wti{T}_{1/1}^{\mu}$ from $T_{1/1}^{\mu}$ if we replace
\begin{equation*}
  \begin{split}
    c_i,\ i & \in\{1,2,5,6,7,8,9,12,13,14,15,16,17,21,22,23,24,25,26,27,28,29, \\
            & 33,34,35,36,37,84,85,86,87,88,89,90,91,92,110,113,114\} \\
  \end{split}
\end{equation*}
by
\begin{equation*}
  \begin{split}
    \ti{c}_1= & \ c_1+2\,c_3+c_4,\ \ti{c}_2=c_2+c_4,\ \ti{c}_5=c_5-c_4,
                \ \ti{c}_6=c_6-c_4, \\
    \ti{c}_7= & \ c_7-2\,c_3,\ \ti{c}_8=c_8+2\,c_{10}+c_{11},
                \ \ti{c}_9=c_9+c_{11},\ \ti{c}_{12}=c_{12}-c_{11}, \\
    \ti{c}_{13}= & \ c_{13}-c_{11},\ \ti{c}_{14}=c_{14}-2\,c_{10},
                   \ \ti{c}_{15}=c_{15}+c_{18}+c_{19},\ \ti{c}_{16}=
                   c_{16}+c_{18}+c_{20}, \\
    \ti{c}_{17}= & \ c_{17}+c_{19}+c_{20},\ \ti{c}_{21}=c_{21}-c_{19},
                   \ \ti{c}_{22}=c_{22}-c_{20},\ \ti{c}_{23}=c_{23}-c_{20}, \\
    \ti{c}_{24}= & \ c_{24}-c_{18},\ \ti{c}_{25}=c_{25}-c_{18},\ \ti{c}_{26}=
                   c_{26}-c_{19},\ \ti{c}_{27}=c_{27}+c_{30}+c_{31}, \\
    \ti{c}_{28}= & \ c_{28}+c_{30}+c_{32},\ \ti{c}_{29}=c_{29}+c_{31}+c_{32},
                   \ \ti{c}_{33}=c_{33}-c_{31},\ \ti{c}_{34}=c_{34}-c_{32}, \\
    \ti{c}_{35}= & \ c_{35}-c_{32},\ \ti{c}_{36}=c_{36}-c_{30},\ \ti{c}_{37}=
                   c_{37}-c_{30},\ \ti{c}_{38}=c_{38}-c_{31}, \\
    \ti{c}_{84}= & \ c_{84}+c_{93}+c_{94},\ \ti{c}_{85}=c_{85}+c_{93}+c_{95},
                   \ \ti{c}_{86}=c_{86}+c_{94}+c_{95},\ \ti{c}_{87}=c_{87}
                   -c_{94}, \\
    \ti{c}_{88}= & \ c_{88}-c_{95},\ \ti{c}_{89}=c_{89}-c_{95},\ \ti{c}_{90}
                   =c_{90}-c_{93},\ \ti{c}_{91}=c_{91}-c_{93}, \\
    \ti{c}_{92}= & \ c_{92}-c_{94},\ \ti{c}_{110}=c_{110}+c_{111},
                   \ \ti{c}_{113}=c_{113}+c_{111},\ \ti{c}_{114}=c_{114}
                   -c_{111} \\
  \end{split}
\end{equation*} 
In the following we will always use this new $Q$-vertex $\wti{T}_{1/1}^{\mu}$. After elimination of these redundant terms in the types $A,\ldots,D,H$ and $K$ one can express the corresponding terms of $d_QT_1$ in an unique way as a divergence in the sense of vector analysis. This is done in appendix A. For the types $E,F,G$ and $J$ the situation is a little different. Here we have no monomial with two derivatives, one of which acting with respect to $x^{\mu}$. For these types it is impossible to obtain an unique divergence for $d_QT_1$, see appendix B. Nevertheless these types are important for the following, as we will see. Let us look at type $E$ first.

\subsection{Type E divergences}
In this subsection we consider the type $E$ divergences explicitly. From the comparison of these divergences with  $d_QT_1|_{Type E}$ we will get linear relations among the coupling parameters of $T_1$. We require the following equation to be satisfied
\begin{equation}
  d_Q T_1|_{Typ E} = \partial_{\mu} \wti{T}_{1/1}^{\mu,E}
  \label{eq:Eichinvarianz-TypE}
\end{equation}
Calculating the right side of this equation we get
\begin{equation}
  \begin{split}
    \partial_{\mu} \wti{T}_{1/1}^{\mu,E} = & \ d_{29} u^{\al}_{\pmu\psig} h^{\al\nu}_{\pnu} h^{\mu\si}+
                                             d_{30} u^{\al}_{\pmu} h^{\al\nu}_{\psig\pnu} h^{\mu\si}+
                                             d_{31} u^{\al}_{\pmu} h^{\al\nu}_{\pnu} h^{\mu\si}_{\psig}+
                                             d_{32} u^{\al}_{\pmu\psig} h^{\al\nu} h^{\mu\si}_{\pnu} \\
                                         & + d_{33} u^{\al}_{\pmu} h^{\al\nu}_{\psig} h^{\mu\si}_{\pnu}+
                                             d_{34} u^{\al}_{\pmu} h^{\al\nu} h^{\mu\si}_{\psig\pnu}+
                                             d_{35} u^{\al} h^{\al\nu}_{\pmu\psig} h^{\mu\si}_{\pnu}+
                                             d_{36} u^{\al} h^{\al\nu}_{\pmu} h^{\mu\si}_{\psig\pnu} \\
                                         & + d_{37} u^{\al}_{\pmu\pnu} h^{\al\nu}_{\psig} h^{\mu\si}+
                                             d_{38} u^{\al}_{\pnu} h^{\al\nu}_{\psig\pmu} h^{\mu\si}+
                                             d_{39} u^{\al}_{\pnu} h^{\al\nu}_{\pmu} h^{\mu\si}_{\psig}+
                                             d_{40} u^{\al}_{\psig\pnu} h^{\al\nu} h^{\mu\si}_{\pmu} \\
                                         & + d_{41} u^{\al}_{\pnu} h^{\al\nu} h^{\mu\si}_{\pmu\psig}+
                                             d_{42} u^{\al} h^{\al\nu}_{\pmu\pnu} h^{\mu\si}_{\psig}+
                                             d_{43} u^{\al} h^{\al\nu}_{\pnu} h^{\mu\si}_{\pmu\psig}+
                                             d_{44} u^{\al}_{\psig\pnu\pmu} h^{\al\nu}h^{\mu\si} \\
                                         & + d_{45} u^{\al} h^{\al\nu}_{\psig\pnu\pmu} h^{\mu\si}+
                                             d_{46} u^{\al} h^{\al\nu} h^{\mu\si}_{\pmu\pnu\psig} \\
  \end{split}
  \label{eq:TypE-Monome}
\end{equation}
The new constants are defined as follows
\begin{equation}
  \begin{split}
    d_{29} := & \ c_{42}+c_{48},\quad d_{30} := c_{40}+c_{42}+c_{51},\quad d_{31} :=  c_{42}+c_{47}+c_{52} \\
    d_{32} := & \ c_{43}+c_{48},\quad d_{33} := c_{43}+c_{44}+c_{51},\quad d_{34} := c_{41}+c_{43}+c_{52} \\ 
    d_{35} := & \ c_{44}+c_{49},\quad d_{36} := c_{41}+c_{44}+c_{53},\quad d_{37} := c_{39}+c_{45}+c_{51} \\
    d_{38} := & \ c_{45}+c_{49},\quad d_{39} := c_{45}+c_{46}+c_{53},\quad d_{40} := c_{39}+c_{46}+c_{52} \\
    d_{41} := & \ c_{46}+c_{50},\quad d_{42} := c_{40}+c_{47}+c_{53},\quad d_{43} := c_{47}+c_{50} \\ 
    d_{44} := & \ c_{39}+c_{48},\quad d_{45} := c_{40}+c_{49},\quad d_{46} := c_{41}+c_{50} \\
  \end{split}
  \label{eq:d-TypE}
\end{equation}
From equation~(\ref{eq:Eichinvarianz-TypE}) we see that
\begin{equation}
  \begin{split}
    d_{29} = & -\frac{i}{2}\,a_{10},\quad d_{30} = -ib_{19},\quad d_{31} = -i\bigl(a_{11}+b_{14}+b_{20}\bigr) \\
    d_{32} = & -\frac{i}{2}\,a_{9},\quad d_{33} = -ia_{12},\quad d_{34} = -ib_{13} \\
    d_{35} = & \ 0,\quad d_{36} = -ib_{18},\quad d_{37} = -i\bigl(2a_{8}+\frac{1}{2}\,a_{9}\bigr) \\
    d_{38} = & \ 0,\quad d_{39} = -i\bigl(a_{7}+b_{17}\bigr),\quad d_{40} = -i\bigl(a_{7}+\frac{1}{2}\,a_{10}\bigr) \\
    d_{41} = & -ib_{16},\quad d_{42} = -ib_{21},\quad d_{43} = -ib_{15} \\
    d_{44} = & \ 0,\quad d_{45} = 0,\quad d_{46} = 0 \\
  \end{split}
  \label{eq:Zuordnung-TypE}
\end{equation}
Finally we arrive at the divergence form if we invert the system~(\ref{eq:d-TypE}). This is done in appendix B. Let $M_E\in Mat(18\times 15,\mathbb{Z})$ be the coefficient matrix of (\ref{eq:d-TypE}). Then we can write this system of equations as
\begin{equation}
  M_E\cdot \mathbf{c}^E=\mathbf{d}^E
\label{eq:d-TypE-MatrixForm}
\end{equation}
where $\mathbf{c}^E\in\mathbb{C}^{15}$ and $\mathbf{d}^E\in\mathbb{C}^{18}$ are the column vectors with components $(c_{39},\ldots,c_{53})$ and $(d_{29},\ldots,d_{46})$ respectively. Now we observe two things:
\begin{enumerate}
  \item  For a solution to exist it is necessary to fulfil 
    \begin{equation}
      \begin{split}
        d_{32} & +d_{36}-d_{33}-d_{46}-d_{44}-d_{39}+d_{37}+d_{41}= 0 \\
        d_{32} & +d_{36}-d_{33}-2d_{44}+d_{40}-d_{39}+d_{37}-d_{31}+d_{29}+d_{43}-d_{46}= 0 \\
        d_{32} & -d_{33}-2d_{44}+d_{40}-d_{39}+d_{37}+d_{35}-d_{31}+d_{29}+d_{42}-d_{45}= 0 \\
        d_{36} & +2d_{32}-d_{33}-2d_{44}+d_{40}-d_{39}+d_{37}-d_{34}= 0 \\
        d_{42} & +2d_{29}-d_{30}+d_{40}-2d_{44}-d_{39}+d_{37}-d_{31}= 0 \\
        d_{38} & +d_{40}-d_{39}-d_{44}-d_{31}+d_{29}+d_{42}-d_{45}= 0 \\
      \end{split}
      \label{eq:TypE-d-Relationen}
    \end{equation}
  \item $rank(M_E)=12$
\end{enumerate}
 
From 2. we get the information that the representation of $d_QT_1|_{Type E}$ as a divergence is not unique. But the important results are the equations (\ref{eq:TypE-d-Relationen}), because we obtain relations among the coupling parameters if we use (\ref{eq:Zuordnung-TypE}):  
\begin{align}
  a_{7} & -2a_{8}-a_{9}+a_{12}-b_{16}+b_{17}-b_{18}= 0 \label{eq:TypE-Relation1} \\
  -2a_{8} & -a_{9}-a_{10}+a_{11}+a_{12}+b_{14}-b_{15}+b_{17}-b_{18}+b_{20}= 0 \label{eq:TypE-Relation2} \\
  -2a_{8} & -a_{9}-a_{10}+a_{11}+a_{12}+b_{14}+b_{17}+b_{20}-b_{21}= 0 \label{eq:TypE-Relation3} \\
  -2a_{8} & -\frac{3}{2}a_{9}-\frac{1}{2}a_{10}+a_{12}+b_{13}+b_{17}-b_{18}= 0 \label{eq:TypE-Relation4} \\
  -2a_{8} & -\frac{1}{2}a_{9}-\frac{3}{2}a_{10}+a_{11}+b_{14}+b_{17}+b_{19}+b_{20}-b_{21}= 0 \label{eq:TypE-Relation5} \\
  -a_{10} & +a_{11}+b_{14}+b_{17}+b_{20}-b_{21}= 0 \label{eq:TypE-Relation6}
\end{align}
These equations are direct consequences of first order gauge invariance.

\subsection{Divergences of Type $F,G,J$}
In analogy to the case of type $E$ we obtain linear relations among the coupling parameters from the types $F,G$ and $J$. One finds the following $9$ relations
\begin{align}
  -a_4 & -a_5-a_9-a_{10}-b_2+b_3-b_4= 0 \label{eq:Rela7} \\
  -2a_3 & -2a_6-a_9-a_{10}-a_{11}-a_{12}+b_3-2b_4+2b_5-b_6= 0 \label{eq:Rela8} \\
  a_5 & -a_6= 0 \label{eq:Rela9} \\
  -a_4 & +a_5-2a_6-\frac{1}{2}a_9-\frac{1}{2}a_{10}-b_1-b_4= 0 \label{eq:Rela10} \\
  -2a_2 & -a_5+a_6-a_8+b_8+b_{11}= 0 \label{eq:Rela11} \\
 -2a_1 & +2a_2-a_4+2a_5-3a_6-a_7+a_8-b_8-2b_{10}+b_{11}-2b_{12}= 0 \label{eq:Rela12} \\
  a_5 & -a_6-b_9-b_{12}= 0 \label{eq:Rela13} \\
  -\frac{1}{2}a_4 & +\frac{3}{2}a_5-2a_6+b_7-b_8-b_{12}= 0 \label{eq:Rela14} \\
  -2a_2 & +a_5-a_6-a_8-b_8-b_{11}= 0 \label{eq:Rela15}
\end{align}
Together with the six relations from type $E$ (\ref{eq:TypE-Relation1}--\ref{eq:TypE-Relation6}) we get $15$ independent equations which restrict the admissible theories. By construction these equations are necessary for a spin-$2$ theory to be gauge invariant.

\subsection{Nilpotency of $Q$}
The gauge charge operator is by definition nilpotent $(Q^2=0)$. As a consequence the application of twice the gauge variation to every expression must vanish, i.e.
\begin{equation}
  (d_Q)^2T_1(x)=0
  \label{eq:Q2}
\end{equation}
If we now use the gauge invariance of $T_1$ to first order, we get additional constraints for the $Q$-vertex $T_{1/1}^{\mu}$, namely
\begin{equation}
  d_Q\left(\partial_{\mu}^x T_{1/1}^{\mu}(x)\right)=0
  \label{eq:Q2-Bedingung}
\end{equation}
This equation gives us restrictions on the parameters of $T_{1/1}^{\mu}$. After a lengthy calculation one arrives at exactly $63$ linear independent coefficients.
The remaining ones can be expressed as linear combinations of them. One might think that these linear dependences may produce further necessary conditions beside the fifteen above. But this is not the case. The divergence expressions become larger if we restrict ourself to these $63$ independent parameters since terms of different types mix up. In view of this it is more convenient to work with the full $100$ different parameters but type by type separately. The equation (\ref{eq:Q2-Bedingung}) is always satisfied as soon as gauge invariance to first order holds. 

\section{Gauge invariant Spin-$2$ theories}
The preceding section has shown what kind of restrictions we obtain if we require the theory to be gauge invariant. The $15$ equations (\ref{eq:TypE-Relation1}--\ref{eq:Rela15}) we have found for the $33$ parameters $a_1,\ldots,a_{12}$ as well as $b_1,\ldots,b_{21}$ play a central role. Now we can look at an arbitrary solution to this set of equations. The corresponding $T_1$ is then gauge invariant to first order because for the following reason. We have to write the gauge variation of this $T_1$ as a divergence in the sense of vector analysis. Because of the generality of our ansatz for the $Q$-vertex every term in $d_QT_1$ can be uniquely identified with a $d_j$-monomial in $\wti{T}_{1/1}^{\mu}$. With the help of the equations from appendix A we can then find a unique divergence for the types $A,B,C,D,H$ and $K$. For the other types we can also find a divergence but in this case it's no longer unique (see appendix B).

Summing up we have proven the following proposition
\begin{proposition}
Let $T_1$ and $\wti{T}_{1/1}^{\mu}$ be given as above, furthermore let $f$ be the following mapping
\begin{align*}
  f : \text{(Vectorspace of Wick-monomials)} & \longrightarrow (\text{Vectorspace of coefficients}\  a_i,\,b_j) \\
          a_1h^{\mu\nu}_{\pmu}h^{}_{\pnu}h+\ldots+
          b_{21}u^{\mu}\ti{u}^{\mu}_{\pnu}h^{\ro\nu}_{\pro} & \longmapsto (a_1,\ldots,a_{12},b_1,\ldots,b_{21})
\end{align*}
Let $V\in\mathbb{R}^{33}$ be the space of solutions to (\ref{eq:TypE-Relation1}--\ref{eq:Rela15}). $V$ is an $18$-dimensional subspace of $\mathbb{R}^{33}$, which is characterised through the following injective linear mapping $L:\mathbb{R}^{18}\longrightarrow \mathbb{R}^{33}$ :
\begin{equation*}
  \begin{split}
   &  \bigl(a_6,a_{12},b_3,b_4,b_5,b_6,b_7,b_{10},b_{11},b_{12},b_{13},b_{14},b_{16},b_{17},b_{18},b_{19},
    b_{20},b_{21}\bigr)\longmapsto \\
   & \Bigl(-b_7-b_{10}-\frac{1}{2}\bigl(b_{16}-b_{17}+b_{18}\bigr), \frac{1}{4}\bigl(b_{13}+b_{17}-b_{18}
     -b_{19}\bigr),-a_6+\frac{1}{2}b_3-b_4+b_5 \\
   & -\frac{1}{2}\bigl(b_6+3\,b_{13}-b_{14}\bigr)-b_{17}+\frac{1}{2}\bigl(3\,b_{18}-b_{19}+b_{20}-b_{21}\bigr),
     -a_6+2\,\bigl(b_7+b_{11}-b_{12}\bigr), \\
   & a_6,a_6,b_{16}-b_{17}+b_{18},\frac{1}{2}\bigl(-b_{13}-b_{17}+b_{18}+b_{19}\bigr),a_{12}+b_{13}+b_{17}
     -b_{18}-b_{19},-a_{12} \\
   & +b_{13}+b_{17}-b_{18}+b_{19},-a_{12}+b_{13}-b_{14}-b_{18}+b_{19}-b_{20}+b_{21},a_{12},-b_4-2\,\bigl(b_7 \\
   & +b_{11}-b_{12}\bigr)-b_{13}-b_{17}+b_{18},b_3-b_4-2\,\bigl(b_7+b_{11}-b_{12}+b_{13}+b_{17}-b_{18}\bigr), \\
   & b_3,b_4,b_5,b_6,b_7,-b_{11},-b_{12},b_{10},b_{11},b_{12},b_{13},b_{14},-b_{18}+b_{21},b_{16},b_{17},
     b_{18},b_{19},b_{20},b_{21}\Bigr) \\
  \end{split}
\end{equation*}
Then we have the two equivalent statements:
\begin{align*}
  (A1)\quad  & d_Q T_1(x)=\partial_{\mu}^x \wti{T}_{1/1}^{\mu}(x)\quad 
       \text{and}\quad d_Q\left(\partial_{\mu}^x \wti{T}_{1/1}^{\mu}(x)\right)=0 \\
 (A2)\quad  & f(T_1)\in V=\text{im}(L)
\end{align*}
where $\text{im}(L)$ means the image of the linear mapping $L$.
\end{proposition} 
This proposition determines all gauge invariant spin-$2$ theories up to first order of perturbation theory. One obtains the trilinear coupling in the expansion of the E-H action by $L(0,1,-1,-1,\underbrace{0,\ldots,0}_{11\,\text{times}},1,0,0)$. This has the explicit form:
\begin{equation}
  T_1^{h,EH}=\kappa\bigl[-\frac{1}{4}\,h^{\mu\nu}h^{}_{\pmu}h^{}_{\pnu}+\frac{1}{2}\,h^{\mu\nu}h^{\al\be}_{\pmu}h^{\al\be}_{\pnu}
             +h^{\mu\nu}h^{\nu\al}_{\pbe}h^{\mu\be}_{\pal}\bigr]
  \label{eq:Einstein-Hilbert-Kopplung}
\end{equation}
The ghost coupling turns out to be the one first suggested by Kugo and Ojima~\cite{ko:scpsmuimqgt}, namely
\begin{equation}
  T_1^{u,KO}=\kappa\bigl[u^{\ro}_{\pnu}\ti{u}^{\mu}_{\pro}h^{\mu\nu}-u^{\ro}\ti{u}^{\mu}_{\pnu}h^{\mu\nu}_{\pro}-u^{\ro}_{\pro}
             \ti{u}^{\mu}_{\pnu}h^{\mu\nu}+u^{\mu}_{\pnu}\ti{u}^{\mu}_{\pro}h^{\ro\nu}\bigr]
  \label{eq:Kugo-Ojima-Kopplung}
\end{equation}
From the viewpoint of gauge properties of a quantised tensor field we have obtained a set of $18$ linear independent gauge theories. We claim that these $18$ different theories together with all their linear combinations are physically equivalent (in the sense explained below) to the trilinear coupling of Einstein-Hilbert~(\ref{eq:Einstein-Hilbert-Kopplung}) plus the ghost coupling of Kugo-Ojima~(\ref{eq:Kugo-Ojima-Kopplung}) up to first order of perturbation theory. Let $P_{phys}$ be the projection from the hole Fock-space $\mathcal{F}$ onto the physical subspace $\mathcal{F}_{phys}$, which can be expressed in terms of the kernel and the range of the gauge charge operator $Q$ by 
\begin{equation}
  \mathcal{F}_{phys}=\text{ker}\,Q/\text{ran}\,Q
  \label{eq:physikalischer-Unterraum}
\end{equation}
(see e.g.~\cite{kr:cptmvbt,gr:cqg2}). Then two $S$-matrices $S,S^{\prime}$ describe the same physics if all matrix elements between physical states agree in the adiabatic limit $g\rightarrow 1$, i.e.
\begin{equation}
  \lim_{g\rightarrow 1}(\phi,P_{phys}S(g)P_{phys}\psi)=\lim_{g\rightarrow 1}(\phi,P_{phys}S^{\prime}(g)P_{phys}\psi),\ \forall \phi,\psi\in \mathcal{F}
  \label{eq:physikalische-Aequivalenz}
\end{equation}
For theories with massless fields the existence of the adiabatic limit is a problem. To avoid this we work with a perturbative version of (\ref{eq:physikalische-Aequivalenz}):
\begin{equation}
  P_{phys}T_nP_{phys}-P_{phys}T^{\prime}_nP_{phys}=\text{divergences}
  \label{eq:physikalische-Aequivalenz:stoerungstheoretische-Form}
\end{equation}
Obviously (\ref{eq:physikalische-Aequivalenz:stoerungstheoretische-Form}) for all $n$ implies (\ref{eq:physikalische-Aequivalenz}) if the adiabatic limit exists. Specialising to first order $n=1$ we see that two couplings $T_1$ and $T^{\prime}_1$ which differ by a divergence are physically equivalent to first order. Furthermore, if they differ by a coboundary, i.e. a term 
\begin{equation}
  T_1^{cb}=d_QX
  \label{eq:Corand-Theorien}
\end{equation}
where $X$ has ghostnumber $n_g(X)=-1$, they are also equivalent because of the equation
\begin{equation}
  P_{phys}(d_QX)P_{phys}=P_{phys}QXP_{phys}+P_{phys}XQP_{phys}=0
  \label{eq:projizierte-Coraender}
\end{equation}
since by inspection of (\ref{eq:physikalischer-Unterraum}) we have  
\begin{equation}
   QP_{phys}=0=P_{phys}Q
  \label{eq:QPPQ}
\end{equation}

Let us return to the space of solutions $V$ from proposition 1. Every vector in $V$ corresponds through the mapping $f^{-1}$ to a gauge invariant theory to first order of perturbation theory. As was mentioned earlier the trilinear coupling of Einstein-Hilbert lies in this space. We now look at the other theories beside the E-H coupling. For this purpose we choose a suitable basis in $V$. It turns out that a basis can be choosen which shows that all theories beside the classical E-H coupling consists of divergences and coboundaries only. Then we have the following theorem. 
\begin{proposition}
Up to first order of perturbation theory all gauge invariant trilinear self-couplings of a quantised tensor field $h^{\mu\nu}(x)$ are physically equivalent to the one obtained from the expansion of the Einstein-Hilbert lagrangian (given by~(\ref{eq:Einstein-Hilbert-Kopplung}) without the two divergence terms, see~\cite{sch:giqgca}).
\end{proposition}
The proof of this proposition is given in appendix C.

Now there arise two questions: 1) Will the statement of this proposition remain true in higher orders? To answer this question we have to show that in each order $n$ we can achieve the form
\begin{equation}
  T_n=T_n^{EH}+d_Q(X_n)+\text{divergences}
\end{equation}
where $T_n^{EH}$ will be constructed from $T_{j}^{EH},\,j=1,\ldots,n-1$ only. We are quite sure that this is indeed the case so the divergence- or coboundary contributions will have no physical effect. This will be further investigated in a forthcomming paper. 

2) What about the gauge invariance of the Einstein-Hilbert coupling in higher orders? In~\cite{sch:giqgca} Schorn obtained the result that the E-H coupling in combination with the Kugo-Ojima-coupling for the ghosts is gauge invariant to second order. There it was necessary to introduce normalisation terms which coincide with the four graviton coupling obtained from the expansion of the E-H lagrangian. Higher than second order have not been investigated up to now. 

\section{Discussion and Outlook}
In this work we have given a detailed analysis of the gauge properties of a quantised tensor field. Very strong restrictions on the admissible form of the interaction are obtained through the requirement of perturbative gauge invariance even in first order of perturbation theory. Among all solutions to our set of equations only the E-H coupling remains as a physically relevant theory. This fact is very remarkable since in our approach only the gauge properties of a quantum field describing a spin-$2$ particle were considered and no use was made of any geometrical input from classical general relativity. In view of this and with the preceding work about Yang-Mills theories in mind we have seen that the principle of operator gauge invariance is a really universal.

In the future we will analyze the Einstein-Hilbert coupling in higher orders of perturbation theory. First of all we will work out a proof of proposition 2 in higher orders which seems possible to us without to many difficulties. Then we are interested in a detailed analysis of the second order gauge invariance for the E-H coupling. Although this was already done by Schorn who has found that the second order is indeed gauge invariant, we hope that we can give a more straightforward proof of this result which can be generalized to higher orders. We also plan to consider other non-flat backgrounds.       
\appendix

In the subsequent appendices A and B we give the explicit divergence forms for the various types of $d_QT_1$. The proof of proposition 2 is outlined in appendix C.

\section{Divergences for types $A,B,C,D,H$ and $K$}
Here we give the unique divergence expressions for $d_QT_1$.

1.) \underline{Type $A$:}
We calculate the expression $\partial_{\mu}\wti{T}_{1/1}^{\mu,A}$ explicitely:
\begin{equation}
  \begin{split}
    \partial_{\mu}\wti{T}_{1/1}^{\mu,A} = & \ d_1\, u^{\mu}_{\pal\pmu}h^{\ro\si}_{\pal}h^{\ro\si}+
                                              d_2\, u^{\mu}_{\pal} h^{\ro\si}_{\pmu\pal}h^{\ro\si}+
                                              d_3\, u^{\mu}_{\pal}h^{\ro\si}_{\pmu}h^{\ro\si}_{\pal} \\
                                          & + d_4\, u^{\mu}_{\pmu}h^{\ro\si}_{\pal}h^{\ro\si}_{\pal}+
                                              d_5\, u^{\mu}h^{\ro\si}_{\pmu\pal}h^{\ro\si}_{\pal} \\
  \end{split}
  \label{eq:TypA-Monome}
\end{equation}
The constants $d_1,\ldots,d_5$ are given by
\begin{equation}
  \begin{split}
    d_1 := & \ \ti{c}_1+\ti{c}_5,\quad d_2 :=\ti{c}_1+\ti{c}_7,\quad d_3 :=\ti{c}_1+\ti{c}_6+\ti{c}_7, \\
    d_4 := & \ \ti{c}_2+\ti{c}_5,\quad d_5 :=2\,\ti{c}_2+\ti{c}_6 \\
  \end{split}
  \label{eq:d-TypA}
\end{equation}
From first order gauge invariance we obtain
\begin{equation}
  d_1=0,\quad d_2=0,\quad d_3=-ia_8,\quad d_4=\frac{i}{2}\,a_8,\quad d_5=0
  \label{eq:Zuordnung-TypA}
\end{equation}
The coefficient matrix $M_A$ of (\ref{eq:d-TypA}) is in $GL(5,\mathbb{Z})$. We invert these equations and obtain
\begin{align}
  \ti{c}_1 = & \ d_1+\frac{1}{2}\,\bigl(d_2-d_3+d_5-2\,d_4\bigr) \\
  \ti{c}_2 = & \ \frac{1}{2}\,\bigl(d_5-d_3+d_2\bigr) \\
  \ti{c}_5 = & \ \frac{1}{2}\,\bigl(d_3-d_2-d_5+2\,d_4\bigr) \\
  \ti{c}_6 = & \ d_3-d_2 \\
  \ti{c}_7 = & \ d_4-d_1+\frac{1}{2}\,\bigl(d_2+d_3-d_5\bigr)
\end{align}
These equations give, together with (\ref{eq:Zuordnung-TypA}), the desired divergence for $d_QT_1|_{Type A}$.

2.) \underline{Type $B$:}
We calculate the expression $\partial_{\mu}\wti{T}_{1/1}^{\mu, B}$ explicitely:
\begin{equation}
  \begin{split}
    \partial_{\mu}\wti{T}_{1/1}^{\mu,B} = & \ d_6\, u^{\mu}_{\pmu\pal}h^{}_{\pal}h+ 
                                              d_7\, u^{\mu}_{\pal}h^{}_{\pmu\pal}h+ 
                                              d_8\, u^{\mu}_{\pal}h^{}_{\pmu}h_{\pal} \\
                                          & + d_9\, u^{\mu}_{\pmu}h^{}_{\pal}h^{}_{\pal}+ 
                                              d_{10}\, u^{\mu}h_{\pmu\pal}h_{\pal} \\
  \end{split}
  \label{eq:TypB-Monome}
\end{equation}
The constants $d_6,\ldots,d_{10}$ are given by
\begin{equation}
  \begin{split} 
    d_6 := & \ \ti{c}_8+\ti{c}_{12},\quad d_7:=\ti{c}_8+\ti{c}_{14},\quad d_8:=\ti{c}_8+\ti{c}_{13}+\ti{c}_{14}, \\
    d_9 := & \ \ti{c}_9+\ti{c}_{12},\quad d_{10}:=2\,\ti{c}_9+\ti{c}_{13} \\
  \end{split}
  \label{eq:d-TypB}
\end{equation}
From first order gauge invariance we obtain
\begin{equation}
  d_6=0,\quad d_7=0,\quad d_8=-ia_2,\quad d_9=\frac{i}{2}\,a_2,\quad d_{10}=0
  \label{eq:Zuordnung-TypB}
\end{equation}
The coefficient matrix $M_B$ of (\ref{eq:d-TypB}) is in $GL(5,\mathbb{Z})$. We invert these equations and obtain
\begin{align}
  \ti{c}_8 = & \ d_6+\frac{1}{2}\,\bigl(d_{10}-2\,d_9-d_8+d_7\bigr) \\
  \ti{c}_9 = & \ \frac{1}{2}\,\bigl(d_{10}-d_8+d_7\bigr) \\
  \ti{c}_{12} = & \ \frac{1}{2}\,\bigl(d_8+2\,d_9-d_{10}-d_7\bigr) \\
  \ti{c}_{13} = & \ d_8-d_7 \\
  \ti{c}_{14} = & \ d_9-d_6-\frac{1}{2}\,\bigl(d_{10}-d_8-d_7\bigr)
\end{align}
These equations give, together with (\ref{eq:Zuordnung-TypB}), the desired divergence for $d_QT_1|_{Type B}$.

3.) \underline{Type $C$:}
We calculate the expression $\partial_{\mu}\wti{T}_{1/1}^{\mu, C}$ explicitely:
\begin{equation}
  \begin{split}
    \partial_{\mu}\wti{T}_{1/1}^{\mu,C} = & \ d_{11}\, u^{\al}_{\pnu\pmu}h^{\al\mu}_{\pnu}h+
                                              d_{12}\, u^{\al}_{\pnu}h^{\al\mu}_{\pnu\pmu}h+
                                              d_{13}\, u^{\al}_{\pnu}h^{\al\mu}_{\pnu}h^{}_{\pmu} \\
                                          & + d_{14}\, u^{\al}_{\pnu\pmu}h^{\al\mu}h^{}_{\pnu}+
                                              d_{15}\, u^{\al}_{\pnu}h^{\al\mu}_{\pmu}h^{}_{\pnu}+
                                              d_{16}\, u^{\al}_{\pnu}h^{\al\mu}h^{}_{\pmu\pnu} \\
                                          & + d_{17}\, u^{\al}_{\pmu}h^{\al\mu}_{\pnu}h^{}_{\pnu}+
                                              d_{18}\, u^{\al} h^{\al\mu}_{\pmu\pnu}h^{}_{\pnu}+
                                              d_{19}\, u^{\al} h^{\al\mu}_{\pnu}h^{}_{\pmu\pnu} \\
  \end{split}
  \label{eq:TypC-Monome}
\end{equation}
The constants $d_{11},\ldots,d_{19}$ are given by
\begin{equation}
  \begin{split} 
    d_{11} := & \ \ti{c}_{15}+\ti{c}_{21},\quad d_{12}:=\ti{c}_{15}+\ti{c}_{24},\quad d_{13}:=\ti{c}_{15}+
                \ti{c}_{25}+\ti{c}_{26}, \\
    d_{14} := & \ \ti{c}_{16}+\ti{c}_{22},\quad d_{15}:=\ti{c}_{16}+\ti{c}_{23}+\ti{c}_{24},\quad d_{16}:=
                \ti{c}_{16}+\ti{c}_{25} \\
    d_{17} := & \ \ti{c}_{17}+\ti{c}_{21}+\ti{c}_{22},\quad d_{18}:=\ti{c}_{17}+\ti{c}_{23},\quad d_{19}:=
                \ti{c}_{17}+\ti{c}_{26} \\
  \end{split}
  \label{eq:d-TypC}
\end{equation}
From first order gauge invariance we get
\begin{equation}
  \begin{split}
    d_{11} = & -ia_6,\quad d_{12}=0,\quad d_{13}=-\frac{i}{2}\,a_5,\quad d_{14}=-\frac{i}{2}\,a_5, \\
    d_{15} = & -\frac{i}{2}\,a_4,\quad d_{16}=0,\quad d_{17}=0,\quad d_{18}=0,\quad d_{19}=0 \\
  \end{split}
  \label{eq:Zuordnung-TypC}
\end{equation}
The coefficient matrix $M_C$ of (\ref{eq:d-TypC}) is in $GL(9,\mathbb{Z})$. We invert these equations and obtain
\begin{align}
  \ti{c}_{15} = & \ \frac{1}{2}\,\bigl(d_{11}+d_{12}+d_{14}-d_{15}-d_{17}+d_{18}\bigr) \\  
  \ti{c}_{16} = & \ \frac{1}{2}\,\bigl(d_{11}-d_{13}+d_{14}+d_{16}-d_{17}+d_{19}\bigr) \\
  \ti{c}_{17} = & \ \frac{1}{2}\,\bigl(d_{12}-d_{13}-d_{15}+d_{16}+d_{18}+d_{19}\bigr) \\
  \ti{c}_{21} = & \ \frac{1}{2}\,\bigl(d_{11}-d_{12}-d_{14}+d_{15}+d_{17}-d_{18}\bigr) \\
  \ti{c}_{22} = & \ \frac{1}{2}\,\bigl(-d_{11}+d_{13}+d_{14}-d_{16}+d_{17}-d_{19}\bigr) \\
  \ti{c}_{23} = & \ \frac{1}{2}\,\bigl(-d_{12}+d_{13}+d_{15}-d_{16}+d_{18}-d_{19}\bigr) \\
  \ti{c}_{24} = & \ \frac{1}{2}\,\bigl(-d_{11}+d_{12}-d_{14}+d_{15}+d_{17}-d_{18}\bigr) \\
  \ti{c}_{25} = & \ \frac{1}{2}\,\bigl(-d_{11}+d_{13}-d_{14}+d_{16}+d_{17}-d_{19}\bigr) \\
  \ti{c}_{26} = & \ \frac{1}{2}\,\bigl(-d_{12}+d_{13}+d_{15}-d_{16}-d_{18}+d_{19}\bigr)
\end{align}
These equations give, together with (\ref{eq:Zuordnung-TypC}), the desired divergence for $d_QT_1|_{Type C}$.

4.) \underline{Type $D$:}
We calculate the expression $\partial_{\mu}\wti{T}_{1/1}^{\mu, D}$ explicitely:
\begin{equation}
  \begin{split}
    \partial_{\mu}\wti{T}_{1/1}^{\mu,D} = & \ d_{20}\, u^{\al}_{\pmu\pnu}h^{\al\si}_{\pnu}h^{\si\mu}+
                                              d_{21}\, u^{\al}_{\pnu}h^{\al\si}_{\pmu\pnu}h^{\si\mu}+
                                              d_{22}\, u^{\al}_{\pnu}h^{\al\si}_{\pnu}h^{\si\mu}_{\pmu} \\
                                          & + d_{23}\, u^{\al}_{\pmu\pnu}h^{\al\si}h^{\si\mu}_{\pnu}+
                                              d_{24}\, u^{\al}_{\pnu}h^{\al\si}_{\pmu}h^{\si\mu}_{\pnu}+
                                              d_{25}\, u^{\al}_{\pnu}h^{\al\si}h^{\si\mu}_{\pmu\pnu} \\
                                          & + d_{26}\, u^{\al}_{\pmu}h^{\al\si}_{\pnu}h^{\si\mu}_{\pnu}+
                                              d_{27}\, u^{\al}h^{\al\si}_{\pmu\pnu}h^{\si\mu}_{\pnu}+
                                              d_{28}\, u^{\al}h^{\al\si}_{\pnu}h^{\si\mu}_{\pmu\pnu} \\
  \end{split}
  \label{eq:TypD-Monome}
\end{equation}
The constants $d_{20},\ldots,d_{28}$ are given by
\begin{equation}
  \begin{split}
    d_{20} := & \ \ti{c}_{27}+\ti{c}_{33},\quad d_{21}:=\ti{c}_{27}+\ti{c}_{36},\quad d_{22}:=\ti{c}_{27}+
                \ti{c}_{37}+\ti{c}_{38} \\
    d_{23} := & \ \ti{c}_{28}+\ti{c}_{34},\quad d_{24}:=\ti{c}_{28}+\ti{c}_{35}+\ti{c}_{36},\quad d_{25}:=
                \ti{c}_{28}+\ti{c}_{37}, \\
    d_{26} := & \ \ti{c}_{29}+\ti{c}_{33}+\ti{c}_{34},\quad d_{27}:=\ti{c}_{29}+\ti{c}_{35},\quad d_{28}:=
                \ti{c}_{29}+\ti{c}_{38} \\
  \end{split}
  \label{eq:d-TypD}
\end{equation}
From first order gauge invariance we obtain
\begin{equation}
  \begin{split}
    d_{20} = & -\frac{i}{2}\,a_9,\quad d_{21}=0,\quad d_{22}=-\frac{i}{2}\,a_{10},\quad d_{23}=-ia_{12} \\
    d_{24} = & -\frac{i}{2}\,a_9,\quad d_{25}=0,\quad d_{26}=0,\quad d_{27}=0,\quad d_{28}=0 \\
  \end{split}
  \label{eq:Zuordnung-TypD}
\end{equation}
The coefficient matrix $M_D$ of (\ref{eq:d-TypD}) is in $GL(9,\mathbb{Z})$. We invert these equations and obtain
\begin{align}
  \ti{c}_{27} = & \ \frac{1}{2}\,\bigl(d_{20}+d_{21}+d_{23}-d_{24}-d_{26}+d_{27}\bigr) \\
  \ti{c}_{28} = & \ \frac{1}{2}\,\bigl(d_{20}-d_{22}+d_{23}+d_{25}-d_{26}+d_{28}\bigr) \\
  \ti{c}_{29} = & \ \frac{1}{2}\,\bigl(d_{21}-d_{22}-d_{24}+d_{25}+d_{27}+d_{28}\bigr) \\
  \ti{c}_{33} = & \ \frac{1}{2}\,\bigl(d_{20}-d_{21}-d_{23}+d_{24}+d_{26}-d_{27}\bigr) \\
  \ti{c}_{34} = & \ \frac{1}{2}\,\bigl(-d_{20}+d_{22}+d_{23}-d_{25}+d_{26}-d_{28}\bigr) \\
  \ti{c}_{35} = & \ \frac{1}{2}\,\bigl(-d_{21}+d_{22}+d_{24}-d_{25}+d_{27}-d_{28}\bigr) \\
  \ti{c}_{36} = & \ \frac{1}{2}\,\bigl(-d_{20}+d_{21}-d_{23}+d_{24}+d_{26}-d_{27}\bigr) \\
  \ti{c}_{37} = & \ \frac{1}{2}\,\bigl(-d_{20}+d_{22}-d_{23}+d_{25}+d_{26}-d_{28}\bigr) \\
  \ti{c}_{38} = & \ \frac{1}{2}\,\bigl(-d_{21}+d_{22}+d_{24}-d_{25}-d_{27}+d_{28}\bigr)
\end{align}                   
These equations give, together with (\ref{eq:Zuordnung-TypD}), the desired divergence for $d_QT_1|_{Type D}$.

5.) \underline{Type $H$:}
We calculate the expression $\partial_{\mu}\wti{T}_{1/1}^{\mu, H}$ explicitely:
\begin{equation}
  \begin{split}
    \partial_{\mu}\wti{T}_{1/1}^{\mu,H} = & \ d_{80}\, u^{\mu}_{\pnu\pmu}\ti{u}^{\al}_{\pnu}u^{\al}+
                                              d_{81}\, u^{\mu}_{\pnu}\ti{u}^{\al}_{\pmu\pnu}u^{\al}+
                                              d_{82}\, u^{\mu}_{\pnu}\ti{u}^{\al}_{\pnu}u^{\al}_{\pmu} \\
                                          & + d_{83}\, u^{\mu}_{\pnu\pmu}\ti{u}^{\al}u^{\al}_{\pnu}+
                                              d_{84}\, u^{\mu}_{\pnu}\ti{u}^{\al}_{\pmu}u^{\al}_{\pnu}+
                                              d_{85}\, u^{\mu}_{\pnu}\ti{u}^{\al}u^{\al}_{\pmu\pnu} \\
                                          & + d_{86}\, u^{\mu}_{\pmu}\ti{u}^{\al}_{\pnu}u^{\al}_{\pnu}+
                                              d_{87}\, u^{\mu}\ti{u}^{\al}_{\pnu\pmu}u^{\al}_{\pnu}+
                                              d_{88}\, u^{\mu}\ti{u}^{\al}_{\pnu}u^{\al}_{\pmu\pnu} \\
  \end{split}
  \label{eq:TypH-Monome}
\end{equation}
The constants $d_{80},\ldots,d_{88}$ are given by
\begin{equation}
  \begin{split}
    d_{80} := & \ \ti{c}_{84}+\ti{c}_{87},\quad d_{81}:=\ti{c}_{84}+\ti{c}_{90},\quad d_{82}:=\ti{c}_{84}+
                \ti{c}_{91}+\ti{c}_{92}, \\
    d_{83} := & \ \ti{c}_{85}+\ti{c}_{88},\quad d_{84}:=\ti{c}_{85}+\ti{c}_{89}+\ti{c}_{90},\quad d_{85}:=
                \ti{c}_{85}+
                \ti{c}_{91}, \\
    d_{86} := & \ \ti{c}_{86}+\ti{c}_{87}+\ti{c}_{88},\quad d_{87}:=\ti{c}_{86}+\ti{c}_{89},\quad d_{88}:=
                \ti{c}_{86}+\ti{c}_{92} \\
  \end{split}
  \label{eq:d-TypH}
\end{equation}
From first order gauge invariance we obtain
\begin{equation}
  \begin{split}
    d_{80} = & 0,\quad d_{81}=0,\quad d_{82}=\frac{i}{2}\,b_{19},\quad d_{83}=0,\quad  d_{84}=
               -\frac{i}{2}\,\bigl(b_1-b_{19}\bigr), \\ 
    d_{85} = & -\frac{i}{2}\,b_2,\quad d_{86}=-\frac{i}{2}\,\bigl(b_4+b_{19}\bigr),\quad d_{87}=0,\quad 
               d_{88}=-\frac{i}{2}\,b_3 \\
  \end{split}
  \label{eq:Zuordnung-TypH}
\end{equation}
The coefficient matrix $M_H$ of (\ref{eq:d-TypH}) is in $GL(9,\mathbb{Z})$. We invert these equations and obtain
\begin{align}
  \ti{c}_{84} = & \ \frac{1}{2}\,\bigl(d_{80}+d_{81}+d_{83}-d_{84}-d_{86}+d_{87}\bigr) \\
  \ti{c}_{85} = & \ \frac{1}{2}\,\bigl(d_{80}-d_{82}+d_{83}+d_{85}-d_{86}+d_{88}\bigr) \\
  \ti{c}_{86} = & \ \frac{1}{2}\,\bigl(d_{81}-d_{82}-d_{84}+d_{85}+d_{87}+d_{88}\bigr) \\
  \ti{c}_{87} = & \ \frac{1}{2}\,\bigl(d_{80}-d_{81}-d_{83}+d_{84}+d_{86}-d_{87}\bigr) \\
  \ti{c}_{88} = & \ \frac{1}{2}\,\bigl(-d_{80}+d_{82}+d_{83}-d_{85}+d_{86}-d_{88}\bigr) \\
  \ti{c}_{89} = & \ \frac{1}{2}\,\bigl(-d_{81}+d_{82}+d_{84}-d_{85}+d_{87}-d_{88}\bigr) \\
  \ti{c}_{90} = & \ \frac{1}{2}\,\bigl(-d_{80}+d_{81}-d_{83}+d_{84}+d_{86}-d_{87}\bigr) \\
  \ti{c}_{91} = & \ \frac{1}{2}\,\bigl(-d_{80}+d_{82}-d_{83}+d_{85}+d_{86}-d_{88}\bigr) \\
  \ti{c}_{92} = & \ \frac{1}{2}\,\bigl(-d_{81}+d_{82}+d_{84}-d_{85}-d_{87}+d_{88}\bigr)
\end{align}
These equations give, together with (\ref{eq:Zuordnung-TypH}), the desired divergence for $d_QT_1|_{Type H}$.
 
6.) \underline{Type $K$:}
We calculate the expression $\partial_{\mu}\wti{T}_{1/1}^{\mu, K}$ explicitely:
\begin{equation}
  \partial_{\mu}\wti{T}_{1/1}^{\mu,K}=d_{102}\, u^{\si}_{\pmu\pnu}\ti{u}^{\mu}_{\pnu}u^{\si}+
                                      d_{103}\, u^{\si}_{\pnu}\ti{u}^{\mu}_{\pnu\pmu}u^{\si}+
                                      d_{104}\, u^{\si}_{\pmu}\ti{u}^{\mu}_{\pnu}u^{\si}_{\pnu}+
                                      d_{105}\, u^{\si}_{\pal\pmu}\ti{u}^{\al}u^{\si}_{\pmu}
  \label{eq:TypK-Monome}
\end{equation}
The constants $d_{102},\ldots,d_{105}$ are given by
\begin{equation}
  d_{102}:=\ti{c}_{110}+c_{112},\ d_{103}:=\ti{c}_{110}+\ti{c}_{114},\ d_{104}:=-\ti{c}_{110}+c_{112}+
           \ti{c}_{113},\ d_{105}:=\ti{c}_{113}
  \label{eq:d-TypK}
\end{equation}
From first order gauge invariance we obtain
\begin{equation}
  d_{102}=\frac{i}{2}\,b_{18},\quad d_{103}=0,\quad d_{104}=-\frac{i}{2}\,b_{13},\quad d_{105}=\frac{i}{2}\,
  b_{17}
  \label{eq:Zuordnung-TypK}
\end{equation}
The coefficient matrix $M_K$ of (\ref{eq:d-TypK}) is in $GL(4,\mathbb{Z})$. We invert these equations and obtain
\begin{align}
  \ti{c}_{110} = & \ \frac{1}{2}\,\bigl(d_{102}-d_{104}+d_{105}\bigr) \\
  c_{112}      = & \ \frac{1}{2}\,\bigl(d_{102}+d_{104}-d_{105}\bigr) \\
  \ti{c}_{113} = & \ d_{105} \\
  \ti{c}_{114} = & \ d_{103}+\frac{1}{2}\,\bigl(-d_{102}+d_{104}-d_{105}\bigr)
\end{align}
These equations give, together with (\ref{eq:Zuordnung-TypK}), the desired divergence for $d_QT_1|_{Type K}$.

\section{Divergences for Types $E,F,G$ and $J$}
Here we calculate the explicit divergence forms in terms of the coupling parameters $a_1,\ldots,a_{12},b_1,\ldots,b_{21}$ for the types $E,F,G$ and $J$. In contrast to the other types the system of equations between the $c_i$ and $d_j$ are no longer invertible in a unique way. There are some ambiguities, if we express the $c_i$ in terms of the $d_j$. Let us begin with type $E$.

1.) \underline{Type $E$:}
Let $M_E\in Mat(18\times 15,\mathbb{Z})$ the coefficient matrix of the system (\ref{eq:d-TypE}). We have to consider the equation 
\begin{equation}
  M_E\cdot \mathbf{c}^E=\mathbf{d}^E
  \label{eq:d-TypEMatrix}
\end{equation}
The general solution of this equation is the sum of an arbitrary solution and the general solution of the corresponding homogeneous equation
\begin{equation}
  M_E\cdot \mathbf{c}^E=0
  \label{eq:homogene TypE-Matrix}
\end{equation}
The matrix $M_E$ has $rank(M_E)=12$. So there are three free parameters $\lambda_1,\lambda_2,\lambda_3\in\mathbb{C}$ in the solution of (\ref{eq:homogene TypE-Matrix}). We obtain
\begin{equation}
  \begin{split}
    \mathbf{c}^E_0(\la_1,\la_2,\la_3) = & \ \Bigl(-\la_1-\la_2+\la_3,\,-\la_1+\la_2-\la_3,\,\la_1
                                             -\la_2-\la_3,\,-\la_1-\la_2+\la_3,\, \\
                                           & -\la_1-\la_2+\la_3,\,-\la_1+\la_2-\la_3,\,-\la_1+\la_2-\la_3,\,\la_1
                                             -\la_2-\la_3,\, \\
                                           & \ \la_1-\la_2-\la_3,\,\la_1+\la_2-\la_3,\,\la_1-\la_2+\la_3,\,-\la_1
                                             +\la_2+\la_3,\,2\,\la_1,\, \\
                                           & \ 2\,\la_2,\,2\,\la_3\Bigr) \\
  \end{split}
  \label{eq:allgemeine homogene TypE-Loesung}
\end{equation}
A special solution of equation~(\ref{eq:d-TypEMatrix}) is given by
\begin{equation}
  \begin{split}
    \mathbf{c}^E_s = & \ \Bigl(\frac{1}{2}\,\bigl(d_{37}+d_{40}-d_{39}\bigr),\,d_{29}-d_{31}-d_{44}+d_{42}
                          -\frac{1}{2}\,\bigl(d_{39}-d_{37}-d_{40}\bigr),\, \\
                        & \ d_{32}+d_{36}-d_{33}-d_{44}-\frac{1}{2}\,\bigl(d_{39}-d_{37}-d_{40}\bigr),\,d_{29}
                          -d_{44}-\frac{1}{2}\,\bigl(d_{39}-d_{37}-d_{40}\bigr),\, \\
                        & \ d_{32}-d_{44}-\frac{1}{2}\,\bigl(d_{39}-d_{37}-d_{40}\bigr),\,-d_{32}+d_{33}+d_{44}
                          +\frac{1}{2}\,\bigl(d_{39}-d_{37}-d_{40}\bigr),\, \\
                        & \ \frac{1}{2}\,\bigl(d_{39}+d_{37}-d_{40}\bigr),\,\frac{1}{2}\,\bigl(d_{40}+d_{39}
                          -d_{37}\bigr),\,d_{31}-d_{29}+d_{44}+\frac{1}{2}\,\bigl(d_{39}-d_{37} \\
                        & -d_{40}\bigr),\,d_{44}+\frac{1}{2}\,\bigl(d_{39}-d_{37}-d_{40}\bigr),\,d_{44}+d_{31}
                          -d_{29}-d_{42}+d_{45}+\frac{1}{2}\,\bigl(d_{39} \\
                        & -d_{37}-d_{40}\bigr),\,-d_{32}-d_{36}+d_{33}+d_{46}+d_{44}+\frac{1}{2}\,\bigl(d_{39}-d_{37}
                          -d_{40}\bigr),\,0,\,0,\,0\Bigr) \\
  \end{split}
  \label{eq:spezielle inhomogene TypE-Loesung}
\end{equation}
The general solution of (\ref{eq:d-TypEMatrix}) is then given by
\begin{equation}
  \mathbf{c}^E=\mathbf{c}^E_s+\mathbf{c}^E_0
  \label{eq:allgemeine TypE-Loesung}
\end{equation}
With the equations (\ref{eq:Zuordnung-TypE}) we can write the expression $d_QT_1|_{Type E}$ as a divergence.

2.) \underline{Type $F$:}
In analogy to type $E$ we first calculate the expression $\partial_{\mu}\wti{T}_{1/1}^{\mu, F}$. This expression is of the form
\begin{equation}
  \partial_{\mu}\wti{T}_{1/1}^{\mu, F}=\sum_{i=47}^{61}d_i\, \partial_{\mu}\partial_{\ro}\partial_{\si}|
                                       u^{\mu}h^{\ro\nu}h^{\nu\si}
  \label{eq:TypF-Monome}
\end{equation}
Here the three derivatives are distributed among fields in all possible combinations. The new constants $d_i$ are defined as follows
\begin{equation} 
  \begin{split}
    d_{47}:= & \ c_{57}+c_{58}+2\,c_{63},\quad d_{48}:=c_{56}+c_{57}+c_{67},\quad d_{49}:=c_{57}+c_{59}+c_{66}, \\
    d_{50}:= & \ c_{58}+c_{62}+c_{67},\quad d_{51}:=c_{55}+c_{58}+c_{66},\quad d_{52}:=c_{59}+c_{62}+c_{64}, \\
    d_{53}:= & \ c_{55}+c_{59}+2\,c_{65},\quad d_{54}:=c_{54}+c_{66}+c_{60},\quad d_{55}:=c_{60}+c_{61}+c_{64}, \\
    d_{56}:= & \ c_{60}+c_{65},\quad d_{57}:=c_{54}+c_{61}+c_{67},\quad d_{58}:=c_{61}+c_{68}, \\
    d_{59}:= & \ c_{56}+c_{62}+2\,c_{68},\quad d_{60}:=c_{54}+c_{63},\quad d_{61}:=c_{55}+c_{64}+c_{56} \\
  \end{split}
  \label{eq:d-TypF}
\end{equation}
From first order gauge invariance we obtain
\begin{equation}
  \begin{split}
    d_{47} = & -i\bigl(\frac{1}{2}\,a_9+a_{12}\bigr),\quad d_{48}=-ib_1,\quad d_{49}=-\frac{i}{2}\,a_9, \\
    d_{50} = & -i\bigl(\frac{1}{2}\,a_{10}+b_2\bigr),\quad d_{51}=0,\quad d_{52}=-ib_3,\quad d_{53}=0, \\
    d_{54} = & \ i\bigl(a_5+\frac{1}{2}\,a_9+a_{12}\bigr),\quad d_{55}=-ib_4,\quad d_{56}=\frac{i}{2}\,\bigl(2\,a_6
               +a_9+a_{12}\bigr), \\
    d_{57} = & \ \frac{i}{2}\,\bigl(2\,a_4+a_9+a_{10}\bigr),\quad d_{58}=\frac{i}{2}\,\bigl(2\,a_3+a_{10}+a_{11}
               -2\,b_5\bigr), \\
    d_{59} = & -ib_6,\quad d_{60}=0,\quad d_{61}=0 \\
  \end{split}
  \label{eq:Zuordnung-TypF}
\end{equation}
Let $M_F\in Mat(15\times 15,\mathbb{Z})$ the coefficient matrix of (\ref{eq:d-TypF}). Then we determine the general solution of 
\begin{equation}
  M_F\cdot \mathbf{c}^F=\mathbf{d}^F
  \label{eq:d-TypF-MatrixForm}
\end{equation}
where $\mathbf{c}^F\in \mathbb{C}^{15}$ and $\mathbf{d}^F\in \mathbb{C}^{15}$ are the column vectors with components $(c_{54},\ldots,c_{68})$ and $(d_{47},\ldots,d_{61})$ respectively. The matrix $M_F$ has $rank(M_F)=11$. The general solution of the corresponding homogeneous system  
\begin{equation}
  M_F\cdot \mathbf{c}^F=0
  \label{eq:homogene TypF-Matrix}
\end{equation}
is labeled by $4$ independent parameters $\la_1,\ldots,\la_4\in\mathbb{C}$ and is given by
\begin{equation}
  \begin{split}
    \mathbf{c}^F_0(\la_1,\la_2,\la_3,\la_4) = & \ \Bigl(-\la_2+\la_3,\,\la_1+\la_3-\la_4,\,-\la_1-2\,\la_3,\,
                                                   \la_1-\la_2+2\,\la_3,\, \\
                                                 & -\la_1-\la_2,\,-\la_1-\la_3-\la_4,\,-\la_4,\,-\la_3,\,\la_1,\,
                                                   \la_2-\la_3,\,\la_3+\la_4,\, \\
                                                 & \ \la_4,\,\la_2-\la_3+\la_4,\,\la_2,\,\la_3\Bigr) \\
  \end{split}
  \label{eq:allgemeine homogene TypF-Loesung}
\end{equation}
A special solution to (\ref{eq:d-TypF-MatrixForm}) is given by
\begin{equation}
  \begin{split}
    \mathbf{c}^F_s = & \ \Bigl(d_{56}+d_{57}-d_{55}-\frac{1}{2}\,\bigl(d_{53}+d_{59}-d_{52}-d_{61}\bigr),\,
                          \frac{1}{2}\,\bigl(d_{53}-d_{59}-d_{52}+d_{61}\bigr),\, \\
                        & \ d_{59},\,d_{49}+d_{57}-d_{55}-d_{54}-d_{53}-d_{59}+d_{61}+2\,d_{56},\,d_{47}-d_{49}
                          +d_{52} \\
                        & +d_{57}-d_{55}-2\,d_{60}+d_{54},\,\frac{1}{2}\,\bigl(d_{53}+d_{59}+d_{52}-d_{61}\bigr),\,
                          d_{56},\,-d_{56}+d_{55} \\
                        & +\frac{1}{2}\,\bigl(d_{53}+d_{59}-d_{52}-d_{61}\bigr),\,0,\,-d_{56}-d_{57}+d_{55}-d_{60}
                          +\frac{1}{2}\,\bigl(d_{53}+d_{59} \\
                        & -d_{52}-d_{61}\bigr),\,-\frac{1}{2}\,\bigl(d_{53}+d_{59}-d_{52}-d_{61}\bigr),\,0,\,d_{54}
                          -2\,d_{56}-d_{57}+d_{55} \\
                        & +\frac{1}{2}\,\bigl(d_{53}+d_{59}-d_{52}-d_{61}\bigr),\,0,\,0\Bigr) \\ 
  \end{split}
  \label{eq:spezielle inhomogene TypF-Loesung}
\end{equation}
The general solution to (\ref{eq:d-TypF-MatrixForm}) is then given by
\begin{equation}
  \mathbf{c}^F=\mathbf{c}^F_s+\mathbf{c}^F_0
  \label{eq:allgemeine TypF-Loesung}
\end{equation}
With the equations (\ref{eq:Zuordnung-TypF}) we can write the expression $d_QT_1|_{Type F}$ as a divergence.

3.) \underline{Type $G$:}
We calculate $\partial_{\mu}\wti{T}_{1/1}^{\mu, G}$. This has the form
\begin{equation}
  \partial_{\mu}\wti{T}_{1/1}^{\mu, G}=\sum_{i=62}^{79}d_i\, \partial_{\mu}\partial_{\ro}\partial_{\si}|u^{\mu}h^{\ro\si}h
  \label{eq:TypG-Monome}
\end{equation}
The new constants $d_i$ are defined by
\begin{equation}
  \begin{split}
    d_{62}:= & \ c_{72}+c_{78},\quad d_{63}:=c_{70}+c_{72}+c_{81},\quad d_{64}:=c_{72}+c_{77}+c_{82}, \\
    d_{65}:= & \ c_{73}+c_{78},\quad d_{66}:=c_{73}+c_{74}+c_{81},\quad d_{67}:=c_{71}+c_{73}+c_{82}, \\
    d_{68}:= & \ c_{74}+c_{79},\quad d_{69}:=c_{71}+c_{74}+c_{83},\quad d_{70}:=c_{69}+c_{75}+c_{81}, \\
    d_{71}:= & \ c_{75}+c_{79},\quad d_{72}:=c_{75}+c_{76}+c_{83},\quad d_{73}:=c_{69}+c_{76}+c_{82}, \\
    d_{74}:= & \ c_{76}+c_{80},\quad d_{75}:=c_{70}+c_{77}+c_{83},\quad d_{76}:=c_{77}+c_{80}, \\
    d_{77}:= & \ c_{71}+c_{80},\quad d_{78}:=c_{70}+c_{79},\quad d_{79}:=c_{69}+c_{78} \\
  \end{split}
  \label{eq:d-TypG}
\end{equation}
From first order gauge invariance we obtain
\begin{equation}
  \begin{split}
    d_{62} = & -ia_6,\quad d_{63}=-ib_7,\quad d_{64}=-\frac{i}{2}\,a_5,\quad d_{65}=-\frac{i}{2}\,a_5, \\
    d_{66} = & -i\bigl(\frac{1}{2}\,a_4+b_8\bigr),\quad d_{67}=0,\quad d_{68}=-ib_9,\quad d_{69}=0,\\
    d_{70} = & \ i\bigl(a_1+a_6+\frac{1}{2}\,a_7\bigr),\quad d_{71}=-ib_{10},\quad \\
    d_{72} = & \ i\bigl(a_1-b_{11}\bigr)+\frac{i}{2}\,\bigl(a_4+a_5+a_7\bigr),\quad d_{73}=i\bigl(2\,a_2
               +\frac{1}{2}\,a_5+a_8\bigr), \\
    d_{74} = & \ 0,\quad d_{75}=-ib_{12},\quad d_{76}=0,\quad d_{77}=0,\quad d_{78}=0,\quad d_{79}=0 \\
  \end{split}
  \label{eq:Zuordnung-TypG}
\end{equation}
Let $M_G\in Mat(18\times 15,\mathbb{Z})$ be the coefficient matrix of (\ref{eq:d-TypG}). We determine the general solution of 
\begin{equation}
  M_G\cdot \mathbf{c}^G=\mathbf{d}^G
  \label{eq:d-TypG-MatrixForm}
\end{equation}
where $\mathbf{c}^G\in \mathbb{C}^{15}$ and $\mathbf{d}^G\in \mathbb{C}^{18}$ are the column vectors with components $(c_{69},\ldots,c_{83})$ and $(d_{62},\ldots,d_{79})$ respectively. The matrix $M_G$ has $rank(M_G)=12$. The general solution of the corresponding homogeneous system
\begin{equation}
  M_G\cdot \mathbf{c}^G=0
  \label{eq:homogene TypG-Matrix}
\end{equation}
is labeled by three independent parameters $\la_1,\la_2,\la_3\in\mathbb{C}$ and is given by
\begin{equation}
  \begin{split} 
    \mathbf{c}^G_0(\la_1,\la_2,\la_3) = & \ \Bigl(\la_3-\la_1-\la_2,\,\la_2-\la_1-\la_3,\,\la_1-\la_2
                                             -\la_3,\,\la_3-\la_1-\la_2,\,\la_3 \\
                                           & -\la_1-\la_2,\,\la_2-\la_1-\la_3,\,\la_2-\la_1-\la_3,\,\la_1-\la_2
                                             -\la_3,\,\la_1-\la_2 \\
                                           & -\la_3,\,\la_1+\la_2-\la_3,\,\la_1-\la_2+\la_3,\,\la_2+\la_3-\la_1,\,
                                             2\,\la_1,\,2\,\la_2,\,2\,\la_3\Bigr) \\
  \end{split}
  \label{eq:allgemeine homogene TypG-Loesung}
\end{equation}
A special solution to (\ref{eq:d-TypG-MatrixForm}) is given by
\begin{equation}
  \begin{split}
    \mathbf{c}^G_s = & \ \Bigl(\frac{1}{2}\,\bigl(d_{73}-d_{72}+d_{70}\bigr),\,\frac{1}{2}\,\bigl(d_{66}
                          +d_{67}-d_{69}\bigr)-d_{65}-d_{64}+d_{62}+d_{75},\, \\
                        & \ \frac{1}{2}\,\bigl(d_{67}-d_{66}+d_{69}\bigr),\,d_{62}-d_{65}-\frac{1}{2}\,\bigl(d_{69}
                          -d_{66}-d_{67}\bigr),\,\frac{1}{2}\,\bigl(d_{66}+d_{67}-d_{69}\bigr),\, \\
                        & \ \frac{1}{2}\,\bigl(d_{69}+d_{66}-d_{67}\bigr),\,\frac{1}{2}\,\bigl(d_{72}-d_{73}
                          +d_{70}\bigr),\,\frac{1}{2}\,\bigl(d_{72}+d_{73}-d_{70}\bigr),\,d_{65}+d_{64} \\
                        & -d_{62}+\frac{1}{2}\,\bigl(d_{69}-d_{66}-d_{67}\bigr),\,d_{65}+\frac{1}{2}\,\bigl(d_{69}
                          -d_{66}-d_{67}\bigr),\,d_{65}+d_{64}-d_{62} \\
                        & -d_{75}+d_{78}+\frac{1}{2}\,\bigl(d_{69}-d_{67}-d_{66}\bigr),\,d_{74}+\frac{1}{2}\,\bigl(
                          d_{66}-d_{67}-d_{69}\bigr),\,0,\,0,\,0\Bigr) \\
  \end{split}
  \label{eq:spezielle inhomogene TypG-Loesung}
\end{equation}
The general solution to (\ref{eq:d-TypG-MatrixForm}) then reads
\begin{equation}
  \mathbf{c}^G=\mathbf{c}^G_S+\mathbf{c}^G_0
  \label{eq:allgemeine TypG-Loesung}
\end{equation}
With the equations (\ref{eq:Zuordnung-TypG}) we can write the expression $d_QT_1|_{Type G}$ as a divergence.
                     
4.) \underline{Type $J$:}
We calculate the expression $\partial_{\mu}\wti{T}_{1/1}^{\mu, J}$. This has the form
\begin{equation}
  \partial_{\mu}\wti{T}_{1/1}^{\mu, J}=\sum_{i=89}^{101}d_i\, \partial_{\mu}\partial_{\al}\partial_{\ro}|
                                       u^{\mu}\ti{u}^{\al}u^{\ro}
  \label{eq:TypJ-Monome}
\end{equation}
The new constants $d_i$ are defined by
\begin{equation}
  \begin{split}
    d_{89}:= & -c_{96}+c_{102}-c_{103},\quad d_{90}:=-c_{96}+c_{105}-c_{101}+c_{109}, \\
    d_{91}:= & \ c_{96}-c_{104}+c_{108},\quad d_{92}=c_{104}-c_{102},\quad d_{93}:=-c_{97}+c_{108}-c_{100}, \\
    d_{94}:= & \ c_{97}-c_{101}+c_{106}+c_{109},\quad d_{95}:=c_{97}-c_{103}+c_{107}, \\
    d_{96}:= & \ c_{108}-c_{103},\quad d_{97}:=c_{104}-c_{99}-c_{107},\quad d_{98}:=c_{99}-c_{105}+c_{106}, \\
    d_{99}:= & \ c_{107}-c_{100},\quad d_{100}:=c_{99}+c_{100}-c_{102},\quad d_{101}:=c_{98}+c_{101}+c_{109} \\
  \end{split}
  \label{eq:d-TypJ}
\end{equation}
From first order gauge invariance we obtain
\begin{equation}
  \begin{split}
    d_{89} = & -\frac{i}{2}\,\bigl(b_3+b_{18}\bigr),\quad d_{90}=0,\quad d_{91}=-\frac{i}{2}\,\bigl(b_1-b_{13}\bigr), \\
    d_{92} = & \ \frac{i}{2}\,\bigl(b_2+b_{17}\bigr),\quad d_{93}=\frac{i}{2}\,b_{18}+ib_{12},\quad d_{94}=0, \\
    d_{95} = & -\frac{i}{2}\,\bigl(b_1+b_4+2\,b_7+b_{13}\bigr),\quad d_{96}=0,\quad d_{97}=\frac{i}{2}\,b_2+ib_8, \\
    d_{98} = & -\frac{i}{2}\,b_3-ib_9,\quad d_{99}=\frac{i}{2}\,b_{17}+ib_{11},\quad d_{100}=0,\quad d_{101}=0 \\
  \end{split}
  \label{eq:Zuordnung-TypJ}
\end{equation}    
Let $M_J\in Mat(13\times 14,\mathbb{Z})$ be the coefficient matrix of (\ref{eq:d-TypJ}). Then we determine the general solution of
\begin{equation}
  M_J\cdot \mathbf{c}^J=\mathbf{d}^J
  \label{eq:d-TypJ-MatrixForm}
\end{equation}
where $\mathbf{c}^J\in \mathbb{C}^{14}$ and $\mathbf{d}^J\in \mathbb{C}^{13}$ are the column vectors with components $(c_{96},\ldots,c_{109})$ and $(d_{89},\ldots,d_{101})$ respectively. The matrix $M_J$ has $rank(M_J)=9$. The general solution of the corresponding homogeneous system
\begin{equation}
  M_J\cdot \mathbf{c}^J=0
  \label{eq:homogene TypJ-Matrix}
\end{equation}
is labeled by $5$ independent parameters. $\la_1,\ldots,\la_5\in\mathbb{C}$ and is given by
\begin{equation}
  \begin{split}
    \mathbf{c}^J_0(\la_1,\ldots,\la_5) = & \ \Bigl(\la_1-\la_2+\la_3-\la_4,\,\la_4-\la_3,\,-\la_4-\la_2+\la_3
                                              -2\,\la_5,\, \\
                                            & \ \la_1-\la_2,\,\la_3,\,\la_4+\la_2-\la_3+\la_5,\,\la_1-\la_2+\la_3,\,
                                              \la_4,\, \\
                                            & \ \la_1-\la_2+\la_3,\,\la_1,\,\la_2,\,\la_3,\,\la_4,\,\la_5\Bigr)\\
  \end{split}
  \label{eq:allgemeine homogene TypJ-Loesung}
\end{equation}
A special solution to (\ref{eq:d-TypJ-MatrixForm}) is given by
\begin{equation}
  \begin{split}
    \mathbf{c}^J_s = & \ \Bigl(d_{94}-d_{95}-d_{90}+d_{96},\,d_{95}-d_{96},\,d_{101}-d_{95}+d_{94}+d_{96},\, 
                          -d_{97}+d_{96} \\
                        & -d_{91}+d_{94}-d_{95}-d_{90},\,d_{100}+d_{97}+d_{89}-d_{96}+d_{91},\,d_{95}-d_{94}
                          -d_{96},\, \\
                        & \ d_{94}-d_{95}+d_{89}-d_{90},\,-d_{96},\,d_{96}-d_{95}+d_{94}-d_{91}-d_{90},\,0,\,0,\,
                          0,\,0,\,0\Bigr) \\
  \end{split}
  \label{eq:spezielle inhomogene TypJ-Loesung}
\end{equation}
The general solution to (\ref{eq:d-TypJ-MatrixForm}) is then given by
\begin{equation}
  \mathbf{c}^J=\mathbf{c}^J_s+\mathbf{c}^J_0
  \label{eq:allgemeine TypJ-Loesung}
\end{equation}
With the equations (\ref{eq:Zuordnung-TypJ}) we can write the expression $d_QT_1|_{Type J}$ as a divergence.

\section{Proof of Proposition 2}
With the notation of proposition 1 we choose a basis $(v_1,\ldots,v_{17},v_{{\scriptscriptstyle EH}})$ in $V$ which displays the vector $v_{{\scriptscriptstyle EH}}:=L(0,1,-1,-1,\underbrace{0,\ldots,0}_{11\,\text{times}},1,0,0)\in V$ corresponding to the Einstein-Hilbert coupling with Kugo-Ojima ghost coupling explicitly. We can choose the remaining basis vectors $v_1,\ldots,v_{17}$ in such a way that they have the following property:
\begin{equation}
   f^{-1}(v_i)=\sum d_QX+\text{divergences}\ \forall i=1,\ldots,17
  \label{eq:Corand-Divergenz-Summe}
\end{equation}
where $X$ is of the form 
\begin{equation}
  X\sim\partial\,\mid \ti{u}hh\quad \text{or}\quad X\sim\partial\,\mid u\ti{u}\ti{u}
  \label{eq:Coraender}
\end{equation}
We consider the following vectors $v_i,\ i=1,\ldots,17\in V$:
\begin{align}
  v_1    & = (0,0,-1,-1,1,1,\underbrace{0,\ldots,0}_{27\,\text{times}}) \label{eq:v1} \\
  v_2    & = (\underbrace{0,\ldots0}_{8\,\text{times}},1,-1,-1,1,\underbrace{0,\ldots,0}_{21\,\text{times}}) \label{eq:v2} \\
  v_3    & = (\underbrace{0,\ldots,0}_{13\,\text{times}},1,1,0,0,1,\underbrace{0,\ldots,0}_{15\,\text{times}}) \label{eq:v3} \\
  v_4    & = (\underbrace{0,\ldots,0}_{12\,\text{times}},-1,-1,0,1,1,\underbrace{0,\ldots,0}_{16\,\text{times}}) \label{eq:v4} \\
  v_5    & = (\underbrace{0,\ldots,0}_{16\,\text{times}},1,2,\underbrace{0,\ldots,0}_{15\,\text{times}}) \label{eq:v5} \\
  v_6    & = (0,0,1,\underbrace{0,\ldots,0}_{13\,\text{times}},1,\underbrace{0,\ldots,0}_{16\,\text{times}}) \label{eq:v6} \\
  v_7    & = (\underbrace{0,\ldots,0}_{18\,\text{times}},1,1,0,-1,-1,\underbrace{0,\ldots,0}_{10\,\text{times}}) \label{eq:v7} \\
  v_8    & = (0,0,0,2,\underbrace{0,\ldots,0}_{8\,\text{times}},-2,-2,0,0,0,0,1,0,0,-1,\underbrace{0,\ldots,0}_{11\,\text{times}}) \label{eq:v8} \\
  v_9    & = (-1,\underbrace{0,\ldots,0}_{20\,\text{times}},1,\underbrace{0,\ldots,0}_{11\,\text{times}}) \label{eq:v9} \\
  v_{10} & = (\underbrace{0,\ldots,0}_{18\,\text{times}},1,0,-1,-1,0,1,\underbrace{0,\ldots,0}_{9\,\text{times}}) \label{eq:v10} \\
  v_{11} & = (0,0,-3/2,0,0,0,0,0,1,1,1,1,0,-2,-1,-1,\underbrace{0,\ldots,0}_{8\,\text{times}},1,0,0,0,0,0,1,0,0) \label{eq:v11} \\
  v_{12} & = (\underbrace{0,\ldots,0}_{25\,\text{times}},1,1,0,0,0,0,0,1) \label{eq:v12} \\
  v_{13} & = (\underbrace{0,\ldots,0}_{25\,\text{times}},1,0,0,0,0,0,-1,0) \label{eq:v13} \\
  v_{14} & = (0,0,1/2,0,0,0,0,0,-1,1,0,-1,\underbrace{0,\ldots,0}_{19\,\text{times}},1,0) \label{eq:v14} \\
  v_{15} & = (\underbrace{0,\ldots,0}_{24\,\text{times}},-1,0,1,1,0,-1,0,0,0) \label{eq:v15} \\
  v_{16} & = (-1/2,0,0,0,0,0,1,\underbrace{0,\ldots,0}_{20\,\text{times}},1,0,0,0,0,0) \label{eq:v16} 
\end{align}
\begin{equation}
  v_{17} = (\underbrace{0,\ldots,0}_{26\,\text{times}},1,0,-1,-1,0,1,0) \label{eq:v17}
\end{equation}
It's easy to see that these vectors together with $v_{{\scriptscriptstyle EH}}$ form a basis of $V$. What remains to be done is to show that they indeed have the property (\ref{eq:Corand-Divergenz-Summe}). After a lenghty calculation we have found: 
\begin{equation}
  \begin{split}
    f^{-1}(v_1) = & -h^{\al\be}_{\pal}h^{\be\mu}_{\pmu}h-h^{\al\be}_{\pal}h^{\be\mu}h^{}_{\pmu}+h^{\al\be}h^{\be\mu}_{\pal}h^{}_{\pmu}+
                    h^{\al\be}_{\pmu}h^{\be\mu}_{\pal}h \\
                = & \  \partial_{\al}\bigl(h^{\al\be}_{\pmu}h^{\be\mu}h-h^{\al\be}h^{\be\mu}_{\pmu}h\bigr) \\
  \end{split}
\end{equation}
\begin{equation}
  \begin{split}
    f^{-1}(v_2) = & \ h^{\mu\nu}_{\pal}h^{\nu\al}_{\pbe}h^{\mu\be}-h^{\mu\nu}_{\pal}h^{\nu\al}h^{\mu\be}_{\pbe}-h^{\mu\nu}h^{\nu\al}_{\pal}h^{\mu\be}_{\pbe}+
                    h^{\mu\nu}h^{\nu\al}_{\pbe}h^{\mu\be}_{\pal} \\
                = & \  \partial_{\al}\bigl(h^{\mu\nu}h^{\nu\al}_{\pbe}h^{\mu\be}-h^{\mu\nu}h^{\nu\al}h^{\mu\be}_{\pbe}\bigr) \\
  \end{split}
\end{equation}
\begin{equation}
  \begin{split}
    f^{-1}(v_3) = & \ u^{\ro}_{\pnu}\ti{u}^{\mu}h^{\mu\nu}_{\pro}+u^{\ro}\ti{u}^{\mu}_{\pnu}h^{\mu\nu}_{\pro}+u^{\ro}\ti{u}^{\mu}_{\pro}h^{\mu\nu}_{\pnu} \\
                = & \ i\,d_Q\bigl(u^{\mu}\ti{u}^{\nu}_{\pmu}\ti{u}^{\nu}\bigr)+\partial_{\nu}\bigl(u^{\ro}\ti{u}^{\mu}h^{\mu\nu}_{\pro}\bigr) \\
  \end{split}
\end{equation}
\begin{equation}
  \begin{split}
    f^{-1}(v_4) = & -u^{\ro}_{\pnu}\ti{u}^{\mu}_{\pro}h^{\mu\nu}-u^{\ro}_{\pnu}\ti{u}^{\mu}h^{\mu\nu}_{\pro}+u^{\ro}_{\pro}\ti{u}^{\mu}_{\pnu}h^{\mu\nu}+
                    u^{\ro}_{\pro}\ti{u}^{\mu}h^{\mu\nu}_{\pnu} \\
                = & \ \partial_{\nu}\bigl(u^{\ro}_{\pro}\ti{u}^{\mu}h^{\mu\nu}-u^{\nu}_{\pro}\ti{u}^{\mu}h^{\mu\ro}\bigr) \\
  \end{split}
\end{equation}
\begin{equation}
  \begin{split}
    f^{-1}(v_5) = & \ u^{\ro}_{\pro}\ti{u}^{\mu}h^{\mu\nu}_{\pnu}+2\,u^{\ro}\ti{u}^{\mu}_{\pro}h^{\mu\nu}_{\pnu} \\
                = & \  i\,d_Q\bigl(u^{\mu}\ti{u}^{\nu}_{\pmu}\ti{u}^{\nu}\bigr)+\partial_{\nu}\bigl(u^{\ro}_{\pro}\ti{u}^{\mu}h^{\mu\nu}+u^{\ro}\ti{u}^{\mu}_{\pro}
                    h^{\mu\nu} \\
                  & +u^{\ro}\ti{u}^{\mu}h^{\mu\nu}_{\pro}-u^{\nu}_{\pro}\ti{u}^{\mu}h^{\mu\ro}-u^{\nu}\ti{u}^{\mu}_{\pro}h^{\mu\ro}\bigr) \\
  \end{split}
\end{equation}
\begin{equation}
  \begin{split}
    f^{-1}(v_6) = & \ h^{\al\be}_{\pal}h^{\be\mu}_{\pmu}h+u^{\ro}_{\pro}\ti{u}^{\mu}h^{\mu\nu}_{\pnu} \\
                = & -i\,d_Q\bigl(\ti{u}^{\mu}h^{\mu\nu}_{\pnu}h\bigr) \\
  \end{split}
\end{equation}
\begin{equation}
  \begin{split}
    f^{-1}(v_7) = & \ u^{\ro}_{\pmu}\ti{u}^{\mu}_{\pro}h+u^{\ro}_{\pmu}\ti{u}^{\mu}h^{}_{\pro}-u^{\ro}_{\pro}\ti{u}^{\mu}_{\pmu}h-
                    u^{\ro}_{\pro}\ti{u}^{\mu}h^{}_{\pmu} \\
                = & \ i\,d_Q\bigl(\frac{1}{2}\,\ti{u}^{\mu}_{\pmu}hh+\ti{u}^{\mu}h^{}_{\pmu}h\bigr)+\partial_{\mu}\bigl(\frac{1}{2}\,h^{\mu\nu}_{\pnu}hh+
                    u^{\mu}_{\pro}\ti{u}^{\ro}h\bigr) \\
  \end{split}
\end{equation}
\begin{equation}
  \begin{split}
    f^{-1}(v_8) = & \ 2\,h^{\al\be}_{\pal}h^{\be\mu}h^{}_{\pmu}-2\,u^{\ro}_{\pnu}\ti{u}^{\mu}_{\pro}h^{\mu\nu}-2\,u^{\ro}_{\pnu}\ti{u}^{\mu}h^{\mu\nu}_{\pro}
                    +u^{\ro}_{\pmu}\ti{u}^{\mu}_{\pro}h-u^{\ro}_{\pro}\ti{u}^{\mu}_{\pmu}h \\
                = & \ i\,d_Q\bigl(\frac{1}{4}\,\ti{u}^{\mu}_{\pmu}hh+\frac{1}{2}\,\ti{u}^{\mu}h^{}_{\pmu}h+\ti{u}^{\mu}_{\pnu}h^{\mu\nu}h
                    +\ti{u}^{\mu}h^{\mu\nu}_{\pnu}h \\
                  & -\ti{u}^{\mu}h^{\mu\nu}h^{}_{\pnu}\bigr)+\partial_{\mu}\bigl(\frac{1}{4}\,h^{\mu\nu}_{\pnu}hh+h^{\mu\be}h^{\be\al}_{\pal}h+
                    u^{\ro}_{\pro}\ti{u}^{\nu}h^{\nu\mu} \\
                  & -2\,u^{\mu}_{\pnu}\ti{u}^{\ro}h^{\ro\nu}+\frac{1}{2}\,u^{\mu}_{\pro}\ti{u}^{\ro}h-\frac{1}{2}\,u^{\ro}\ti{u}^{\ro}_{\pmu}h+\frac{1}{2}\,
                    u^{\ro}\ti{u}^{\ro}h^{}_{\pmu}\bigr) \\
  \end{split}
\end{equation}
\begin{equation}
  \begin{split}
    f^{-1}(v_9) = & -h^{\mu\nu}_{\pmu}h^{}_{\pnu}h+u^{\ro}_{\pro}\ti{u}^{\mu}_{\pmu}h \\
                = & -\frac{i}{2}\,d_Q\bigl(\ti{u}^{\mu}_{\pmu}hh\bigr)-\frac{1}{2}\,\partial_{\mu}\bigl(h^{\mu\nu}_{\pnu}hh\bigr) \\
  \end{split}
\end{equation}
\begin{equation}
  \begin{split}
    f^{-1}(v_{10}) = & \ u^{\ro}_{\pmu}\ti{u}^{\mu}_{\pro}h-u^{\ro}\ti{u}^{\mu}_{\pmu}h^{}_{\pro}-u^{\ro}_{\pro}\ti{u}^{\mu}_{\pmu}h+
                       u^{\ro}\ti{u}^{\mu}_{\pro}h^{}_{\pmu} \\
                   = & \ \partial_{\mu}\bigl(u^{\ro}\ti{u}^{\mu}_{\pro}h-u^{\mu}\ti{u}^{\ro}_{\pro}h\bigr) \\
  \end{split}
\end{equation}
\begin{equation}
  \begin{split}
    f^{-1}(v_{11}) = & -\frac{3}{2}\,h^{\al\be}_{\pal}h^{\be\mu}_{\pmu}h+h^{\mu\nu}_{\pal}h^{\nu\al}_{\pbe}h^{\mu\be}+h^{\mu\nu}_{\pal}h^{\nu\al}h^{\mu\be}_{\pbe}
                       +h^{\mu\nu}h^{\nu\al}_{\pal}h^{\mu\be}_{\pbe}+h^{\mu\nu}h^{\nu\al}_{\pbe}h^{\mu\be}_{\pal} \\
                     & -2\,u^{\ro}_{\pnu}\ti{u}^{\mu}h^{\mu\nu}_{\pro}-u^{\ro}\ti{u}^{\mu}_{\pnu}h^{\mu\nu}_{\pro}-u^{\ro}_{\pro}\ti{u}^{\mu}_{\pnu}h^{\mu\nu}
                       +u^{\ro}_{\pmu}\ti{u}^{\mu}_{\pnu}h^{\ro\nu}+u^{\mu}_{\pnu}\ti{u}^{\mu}_{\pro}h^{\ro\nu} \\
                   = & \ i\,d_Q\bigl(2\,\ti{u}^{\mu}_{\psig}h^{\mu\nu}h^{\nu\si}+\frac{3}{2}\,\ti{u}^{\mu}h^{\mu\nu}_{\pnu}h-
                       \frac{1}{2}\,u^{\mu}\ti{u}^{\nu}_{\pmu}\ti{u}^{\nu}\bigr)+\partial_{\al}\bigl(h^{\mu\nu}h^{\nu\al}_{\pbe}h^{\mu\be} \\
                     & +h^{\mu\nu}h^{\nu\al}h^{\mu\be}_{\pbe}+\frac{3}{2}\,u^{\ro}_{\pro}\ti{u}^{\mu}h^{\mu\al}+\frac{1}{2}\,u^{\ro}\ti{u}^{\mu}_{\pro}h^{\mu\al}-
                       \frac{1}{2}\,u^{\ro}\ti{u}^{\mu}h^{\mu\al}_{\pro}-\frac{3}{2}\,u^{\al}_{\pnu}\ti{u}^{\mu}h^{\mu\nu} \\
                     & -\frac{1}{2}\,u^{\al}\ti{u}^{\mu}_{\pnu}h^{\mu\nu}-\frac{1}{2}\,u^{\ro}_{\pal}\ti{u}^{\mu}h^{\ro\mu}-
                       \frac{1}{2}\,u^{\ro}\ti{u}^{\mu}_{\pal}h^{\ro\mu}+\frac{1}{2}\,u^{\ro}\ti{u}^{\mu}h^{\ro\mu}_{\pal}\bigr) \\
  \end{split}
\end{equation}
\begin{equation}
  \begin{split}
    f^{-1}(v_{12}) = & \ u^{\ro}_{\pmu}\ti{u}^{\mu}h^{\ro\nu}_{\pnu}+u^{\ro}\ti{u}^{\mu}_{\pmu}h^{\ro\nu}_{\pnu}+u^{\mu}\ti{u}^{\mu}_{\pnu}h^{\ro\nu}_{\pro} \\
                   = & \ i\,d_Q\bigl(u^{\mu}\ti{u}^{\mu}_{\pnu}\ti{u}^{\nu}\bigr)+\partial_{\mu}\bigl(u^{\ro}\ti{u}^{\mu}h^{\ro\nu}_{\pnu}\bigr) \\
  \end{split}
\end{equation}
\begin{equation}
  \begin{split}
    f^{-1}(v_{13}) = & \ u^{\ro}_{\pmu}\ti{u}^{\mu}h^{\ro\nu}_{\pnu}-u^{\mu}_{\pnu}\ti{u}^{\mu}h^{\ro\nu}_{\pro} \\
                   = & -i\,d_Q\bigl(u^{\mu}_{\pnu}\ti{u}^{\mu}\ti{u}^{\nu}\bigr) \\
  \end{split}
\end{equation}
\begin{equation}
  \begin{split}
    f^{-1}(v_{14}) = & \ \frac{1}{2}\,h^{\al\be}_{\pal}h^{\be\mu}_{\pmu}h-h^{\mu\nu}_{\pal}h^{\nu\al}_{\pbe}h^{\mu\be}+h^{\mu\nu}_{\pal}h^{\nu\al}h^{\mu\be}_{\pbe}
                       -h^{\mu\nu}h^{\nu\al}_{\pbe}h^{\mu\be}_{\pal}+u^{\mu}_{\pnu}\ti{u}^{\mu}h^{\ro\nu}_{\pro} \\
                   = & \ i\,d_Q\bigl(\ti{u}^{\mu}h^{\mu\nu}h^{\nu\si}_{\psig}-\frac{1}{2}\,\ti{u}^{\mu}h^{\mu\nu}_{\pnu}h+\frac{1}{2}\,u^{\mu}_{\pnu}\ti{u}^{\mu}
                       \ti{u}^{\nu}\bigr) \\
                     & -\partial_{\al}\bigl(h^{\mu\nu}h^{\nu\al}_{\pbe}h^{\mu\be}-h^{\mu\nu}h^{\nu\al}h^{\mu\be}_{\pbe}\bigr) \\
  \end{split}
\end{equation}
\begin{equation}
  \begin{split}
    f^{-1}(v_{15}) = & -u^{\ro}_{\pmu}\ti{u}^{\mu}_{\pnu}h^{\ro\nu}+u^{\ro}\ti{u}^{\mu}_{\pmu}h^{\ro\nu}_{\pnu}+u^{\ro}_{\pnu}\ti{u}^{\mu}_{\pmu}h^{\ro\nu}-
                       u^{\ro}\ti{u}^{\mu}_{\pnu}h^{\ro\nu}_{\pmu} \\
                   = & \ \partial_{\mu}\bigl(u^{\ro}\ti{u}^{\nu}_{\pnu}h^{\ro\mu}-u^{\ro}\ti{u}^{\mu}_{\pnu}h^{\ro\nu}\bigr) \\
  \end{split}
\end{equation}
\begin{equation}
  \begin{split}
    f^{-1}(v_{16}) = & -\frac{1}{2}\,h^{\mu\nu}_{\pmu}h^{}_{\pnu}h+h^{\mu\nu}_{\pmu}h^{\al\be}_{\pnu}h^{\al\be}+u^{\ro}_{\pnu}\ti{u}^{\mu}_{\pmu}h^{\ro\nu} \\
                   = & \ i\,d_Q\bigl(\frac{1}{2}\,\ti{u}^{\mu}_{\pmu}h^{\al\be}h^{\al\be}-\frac{1}{4}\,\ti{u}^{\mu}_{\pmu}hh\bigr)+\partial_{\mu}\bigl(
                       \frac{1}{2}\,h^{\mu\nu}_{\pnu}h^{\al\be}h^{\al\be}-\frac{1}{4}\,h^{\mu\nu}_{\pnu}hh\bigr) \\
  \end{split}
\end{equation}
\begin{equation}
  \begin{split}
    f^{-1}(v_{17}) = & \ u^{\ro}\ti{u}^{\mu}_{\pmu}h^{\ro\nu}_{\pnu}-u^{\ro}_{\pnu}\ti{u}^{\mu}h^{\ro\nu}_{\pmu}-u^{\ro}\ti{u}^{\mu}_{\pnu}h^{\ro\nu}_{\pmu}
                       +u^{\mu}_{\pnu}\ti{u}^{\mu}h^{\ro\nu}_{\pro} \\
                   = & \ i\,d_Q\bigl(\frac{1}{4}\,\ti{u}^{\mu}_{\pmu}hh+\frac{1}{2}\,\ti{u}^{\mu}h^{}_{\pmu}h-\frac{1}{2}\,\ti{u}^{\mu}_{\pmu}h^{\al\be}
                       h^{\al\be}-\ti{u}^{\mu}h^{\al\be}_{\pmu}h^{\al\be}+u^{\mu}_{\pnu}\ti{u}^{\mu}\ti{u}^{\nu}\bigr) \\
                     & +\partial_{\mu}\bigl(\frac{1}{4}\,h^{\mu\nu}_{\pnu}hh-\frac{1}{2}\,h^{\mu\nu}_{\pnu}h^{\al\be}h^{\al\be}-u^{\ro}\ti{u}^{\mu}_{\pnu}
                       h^{\ro\nu}+u^{\ro}_{\pnu}\ti{u}^{\nu}h^{\ro\mu}+u^{\ro}\ti{u}^{\nu}_{\pnu}h^{\ro\mu}\bigr) \\
  \end{split}
\end{equation}
It should be noted that there is no possibility to write $v_{{\scriptscriptstyle EH}}$ in the form (\ref{eq:Corand-Divergenz-Summe}). Then the theorem is proven because all basis vectors except $v_{{\scriptscriptstyle EH}}$ have a form which lead to unphysical $S$-matrix elements. Together with the discussion preceeding the proposition 2 in section 5 we can now argue that the only physically relevant theory is the coupling of Einstein-Hilbert.


\providecommand{\bysame}{\leavevmode\hbox to3em{\hrulefill}\thinspace}

\end{document}